\documentstyle[12pt,epsfig,epsf,cite]{article}
\def\gsim{\lower0.5ex\hbox{$\:\buildrel >\over\sim\:$}}
\def\lsim{\lower0.5ex\hbox{$\:\buildrel <\over\sim\:$}}
\def\barnue{\:\raisebox{-0.35ex}{$\stackrel{(-)}{\nu_e}$}\:}

\newcommand{\be}{\begin{equation}}
\newcommand{\ee}{\end{equation}}
\newcommand{\beq}{\begin{eqnarray}}
\newcommand{\eeq}{\end{eqnarray}}
\newcommand{\barr}{\begin{array}}
\newcommand{\earr}{\end{array}}
\newcommand{\dis}{\displaystyle}
\newcommand{\ev}{\; {\rm eV}}
\newcommand{\gev}{\; {\rm GeV}}

\def\atil{\widetilde A}

\def\ue{|U_{e 3}|}
\def\Um{|U_{\mu 3}|}
\def\cP{c_{\Psi}}
\def\sP{s_{\Psi}}
\def\cOm{c_{\Omega}}
\def\sOm{s_{\Omega}}

\begin{document}
\begin{flushright}
{ROME1-1391-2004}\\
\end{flushright}

\begin{center}

{\Large\bf Detecting matter effects in long baseline experiments}\\[18mm]
{\bf Debajyoti Choudhury\footnote{On leave from Harish-Chandra Research Institute, Allahabad, India.}  and
Anindya Datta\footnote{E-mail: debchou@physics.du.ac.in, debchou@mail.cern.ch, 
Anindya.Datta@roma1.infn.it} } \\[2ex]

$^1${\em Department of Physics and Astrophysics, University of Delhi,
Delhi 110 007, India.}
\\
$^2${\em INFN, Sezione di Roma; Dip. di
Fisica, Universita La Sapienza, I-00185, Rome, Italy.}
\end{center}

\vskip 20pt
\begin{abstract}
Experiments strongly suggest that the flavour mixing responsible for
the atmospheric neutrino anomaly is very close to being maximal.
Thus, it is of great theoretical as well as experimental importance to
measure any possible deviation from maximality.  In this context, we
reexamine the effects of matter interactions in long baseline neutrino
oscillation experiments.  Contrary to popular belief, the muon
neutrino survival probability is shown to be quite sensitive to matter
effects. Moreover, for moderately long baselines, the difference between
the survival probilities for $\nu_\mu$ and $\bar\nu_\mu$ is shown
to be large and sensitive to the deviation of $|U_{\mu 3}|$ from
maximality. Performing a realistic analysis, we demonstrate that
a muon-storage ring $\nu$-source alongwith an iron calorimeter detector can
measure such deviations. (Contrary to recent claims, this is not so
for the NuMI--{\sc minos} experiment.) We also discuss the possible
correlation in measuring $U_{\mu 3}$ and $U_{e3}$ in such experiment.
\end{abstract}

\section{Introduction}
Neutrino masses and mixings continue to intrigue us, despite
the continued efforts of many recent
experiments~\cite{superK, sno, chooz, kamland, k2k,review}. We do
know that the muon-neutrino ($\nu_\mu$) mixes almost maximally
($\sin^2 2\theta_{\mu3} >  0.92$ at 90\% C.L) with
another species $\nu_3$ (which could be identified with the tau-neutrino,
$\nu_\tau$) and that the mass splitting
$|\delta m^2_{32}| \sim 2 \times 10^{-3} \ev^2$. Furthermore, the
electron-neutrino ($\nu_e$) mixes with a combination of
$\nu_\mu$ and $\nu_\tau$ with a similarly large angle $\theta_{e2} \simeq
30^\circ$ but a far smaller mass splitting
($\delta m^2_{21} \sim 7 \times 10^{-5} \ev^2$). And, finally, the third
mixing angle $\theta_{13}$  is constrained, by the {\sc chooz}
experiment \cite{chooz}, to be small. What we do not know though are
the sign of $\delta m^2_{32}$, the precise value of $\theta_{13}$ and 
the extent to which $\theta_{\mu 3}$ may, if at all, differ from maximality. 
Since such knowledge is obviously essential to the formulation of a
theory of neutrino flavours, the importance of precise determinations 
cannot but be underemphasised. Neutrino oscillations provide 
just the proper platform for such an exercise.

As is well known, during passage through dense matter, various
amplitudes for neutrino oscillations may be magnified to a great
degree \cite{msw} thereby rendering them measurable with relative
ease.  While propagating in a medium, all the three neutrino flavours
interact with matter via neutral curent (NC) interactions with equal
strength. Those of the electron flavour, $\barnue$, have, in addition,
a extra charged current interaction as well.  Since it is this extra
interaction that provides {\em flavour-dependence}, all matter-induced
effects in neutrino oscillations (including those in the
$\nu_\mu$--$\nu_\tau$ sector) are proportional to the size of the
relevant mixing with $\nu_e$.

Recent  theoretical advances in the quest of producing
very high intensity muon sources engender
optimism regarding  a future neutrino factory wherein an intense
beam of muons is to be accelerated to a not too high energy
and stored in a storage ring with a
straight section directed towards a neutrino
detector.  Muons would decay in this straight section thereby producing high
intensity neutrino beams (both of electron-- and muon-types)
that are highly collimated in the direction of the
decaying muons \cite{nu_fac_tech}. The efficacy of muon storage ring
neutrino sources in performing precision measurements of neutrino
parameters has already been discussed in the
literature~\cite{nu_fac}. The advantages over
neutrino experiments
with conventional neutrino beams (arising from $\pi^\pm$ decay)
are manifold: ($i$) a precise knowledge of the $\nu_\mu$ and $\bar
\nu_e$ fluxes helps reduce the systematic
errors; ($ii$) assuming a $\mu$-beam of, say, 
20 GeV energy, such neutrinos have, on the average, energies higher than
those of conventional neutrino beams thereby increasing the $\nu$
cross-section at the detector; and ($iii$) the aforementioned
collimation.

Previous studies of the matter effect in Ref. \cite{barger_matter}  at
a long base line experiment were concentrated mainly on the measurement of
the rate of wrong sign muon events, which, in turn is proportional to
the transition probablity $P_{\nu_e \to \nu_\mu}$. Such a measurement
would enable a determination of (or constraining of) $\sin
\theta_{13}$ {\em provided}\ independent and precise measurements of
the mixing angle $\theta_{23}$ and mass-square difference $\delta m^2
_{31}$ exist. Possible measurements of neutrino mixing parameters from long 
baseline experiments using conventional neutrino beams as well as from 
a neutrino factory, and a water \v{C}erenkov detector has also been discussed 
in Ref.\cite{maoki}.

In the present article, we demonstrate that a sizable
matter effect is also evinced by a measurement of the $\nu_\mu$
survival probability. In addition, such a measurement has the
potential of detecting possible deviations from maximality of the
mixing angle responsible for the atmospheric neutrino anomaly. In a
recent study \cite{raj}, the presence of large matter effect in
$\nu_\mu - \nu_\tau$ oscillation as well as in $\nu_\mu$ survival
probability has been stressed. Although there is a large matter effect
present in $\nu_\mu - \nu_\tau$ sector, measurement of the
same is difficult due to both the lower detection efficiency for
the tau-events as well as the need for a more sophisticated
detector.

Whereas much of our analysis would be source and detector independent,
we also choose to examine the feasibility of making such meaurements
in the context of a realistic experimental setup.  To be concrete, we
choose a 50 kT Iron detector, with detection and charge discrimination
capability for muons provided by a magnetic field. Such a detector was
proposed for Gran Sasso ({\sc monolith}) \cite{mono} and, more
recently, for a location in India (INO) \cite{ino}. The latter is
contemplated primarily as a detector for atmospheric neutrinos and
also as an end detector for a future neutrino factory beam. A muon
detection threshold of 2 GeV \cite{ino} has been used in our
calculation. As for the baseline, we illustrate our results for two
particular choices. The first one coresponds to a distance from JHF to
INO ($\sim$ 5000 Km) and the other one corresponds to even a longer
baseline, from FERMILAB to INO ($\sim$ 10000 Km). We will also compare
results with a baseline of 732 Km, corresponding to the Fermilab-{\sc
minos} distance.

The plan for the rest of the article is as follows.  In section
\ref{sec:analytic}, we discuss neutrino oscillation in the presence of
matter. We also define the asymmetry of $\mu^+$ and $\mu^-$ events and
explain its correlation to $U_{e3}$ and $U_{\mu 3}$.  We present the
expressions for the asymmetry for two different aproximations: when
matter effect is small compared to $\frac{\delta m^2 _{32}}{2 E_\nu}$
i.e for short base line experiments like {\sc minos}
\cite{choubey_roy} and also when the matter term is comparable to the
$\frac{\delta m^2 _{32}}{2 E_\nu}$  i.e., for long base line
experiments. In section \ref{sec:minos}, we briefly review and
reexamine the proposal, in a recent article \cite{choubey_roy}, to
measure matter effect and the above correlations in the context of the
{\sc minos} experiment. Section \ref{sec:longbaseline} deals with the
expectations at lomg baseline experiments, both for the $\nu_\mu$
survival probability as well as for the $\nu_e$ to $\nu_\mu$
transition probability. An analysis of the findings in the context of
a realistic experimental situation is effected in Section
\ref{sec:storage_ring}. Finally, we conclude in Section
\ref{sec:concl}.

\section{Neutrino Oscillation in presence of matter}
      \label{sec:analytic}
We begin by reviewing the passage of neutrino through matter.
While analytical results are possible, the exact answers are
cumbersome and not very useful. However, it is useful to
consider approximations such as a constant density
profile for the earth. While such an
assumption is obviously not a very accurate one, especially when
neutrinos traverse a longer distance through the earth,
the approximation serves to give us some
physical insights to the problem we are dealing with. Furthermore,
results such as those we shall obtain are still applicable if
the matter density within the earth may be approximated to
be piecewise constant. It must be kept in mind though that
the quantitative results presented in this article are
obtained not by means of such an approximation, but by 
explicitly accounting for 
the varying density profile of the earth using the Preliminary
Reference Earth Model (PREM) \cite{prem}.

As with the quark sector, neutrino flavour states can be expressed
in terms of mass eigenstates through a relation of the form
\begin{equation}
\vert \nu_\alpha \rangle \;= \sum_i U_{\alpha i}\;\vert \nu_i \rangle \ ,
\end{equation}
where $U$ is a $3 \times 3$ unitary matrix known as the
Pontecorvo-Maki-Nakagawa-Sakata mixing matrix. In the absence of Majorana
phases (assumed to be so henceforth)\footnote{Any Majorana phase in the
PMNS matrix
cannot be probed by an oscillation experiment.},
$U$ can be parametrized in terms
of three mixing angles and a $CP$-violating phase $\delta_{\rm CP}$,
{\em viz.}
\beq
U = \pmatrix{
          c_{12}c_{13} & s_{12}c_{13} & s_{13}e^{-i\delta_{\rm CP}}  \cr
 -c_{23}s_{12} - s_{23}s_{13}c_{12}e^{i\delta_{\rm CP}} & c_{23}c_{12} -
s_{23}s_{13}s_{12}e^{i\delta_{\rm CP}}&  s_{23}c_{13}\cr
  s_{23}s_{12} - c_{23}s_{13}c_{12}e^{i\delta_{\rm CP}}& -s_{23}c_{12} -
c_{23}s_{13}s_{12}e^{i\delta_{\rm CP}} & c_{23}c_{13} \cr} \ ,
\label{mns}
\eeq
with $c_{ij} \equiv \cos \theta_{ij}$ and $s_{ij} \equiv \sin \theta_{ij}$.
In vacuum, neutrino oscillations are then governed by $U$ and the
two independent differences in the squares of the neutrino masses,
$\delta m^2_{ij} \equiv m_i^2 - m_j^2$. In the rest of our analysis, we will
make a further simplifying assumption of $\delta_{\rm CP} = 0$.
With the above parametrization of $U$, experimental results can
be used to constrain
the mixing angles $\theta_{ij}$ and the mass-square differences.
Measurements of the solar neutrino fluxes, augmented by the
KamLAND reactor data give
\beq
0.30  <  \tan^2 \theta_{12} < 0.64, \qquad
5.4\times 10^{-5} \ev^2  <   \delta_{21} < 9.4 \times 10^{-5} \ev^2
\eeq
with the best-fit values given by $\tan^2 \theta_{12} = 0.40 $
and $\delta m^2_{21} = 6.9 \times 10^{-5} \ev^2$.
%
Experiments with atmospheric neutrinos (as well as the K2K
experiment) suggest, on the other hand, that
\beq
\sin^2 2\theta_{23}  >  0.86, \qquad
1.4 \times 10^{-3} \ev^2 <  |\delta m^2_{32}| < 5.1 \times 10^{-3} \ev^2 \ ,
\eeq
with the best-fit value being $|\delta m^2_{32}| = 2.0 \times
10^{-3}~\rm{eV^2}$.
%
And, finally, the negative result of the {\sc chooz} experiment implies
that
\beq
\sin^2 \theta_{13} < 5\times 10^{-2}~ \qquad {\rm at} \; \hbox{99.73\% C.L.}
\eeq
Note that,
\begin{itemize}
\item unlike in the case for $\delta m^2_{21}$, the sign of $\delta m^2_{32}$ is
undetermined;
\item since $|\delta m^2_{32}| \approx |\delta m^2_{31}| \gg  |\delta m^2_{21}|$
   \footnote{From now on we will denote this large mass-square difference by 
  $\delta m^2_{31}$},
      the last-mentioned (as well as the solar mixing angle $\theta_{12}$)
      plays only a subservient role in $\nu_\mu$ oscillations;
\item while $|U_{\mu 3}|$ is constrained to be close to $1/\sqrt{2}$,
      it is still allowed to be non-maximal.
\end{itemize}

Working in the limit of a vanishingly small $\delta m^2_{21}$,
all the expressions for the oscillation/survival probabilities can be
expressed in terms of just two of the mixing
angles\footnote{While the analytical expressions below
  have been derived under this approximation,
  all of the numerical analysis has been performed
  using the full expression for $U$ and a non-zero value for
  $\delta_{12}$.},
{\em viz.} $\theta_{23}$ and $\theta_{13}$.
For a $\nu_\mu$ ($\bar \nu_\mu$) of
energy $E_\nu$ travelling a distance $L$ in vacuum,
the survival probability is then given by
\beq
P^{\rm vac}_{\mu\mu} = P^{\rm vac}_{\bar \mu \bar \mu} = 1-4 |U_{\mu 3}|^2 (1 -
|U_{\mu 3}|^2) \sin^2 \left(\Delta_{31}L / 2\right) \, ,
\label{survival_vac}
\eeq
where,
\beq
U_{e 3} = s_{13} \ , \qquad U_{\mu 3} = c_{1 3} \, s_{2 3} \ , \qquad
\Delta_{31} \equiv \frac{\delta m^2_{31}}{2 E_\nu} \ .
\eeq
In the
presence of matter, the electron (anti-)neutrino sees an additional potential
corresponding to charged current interactions, and the effective Hamiltonian
is given by
\beq
H = U \pmatrix{0 & 0 & 0 \cr 0 & \Delta_{21} & 0 \cr 0 & 0 & \Delta_{31}}
U^\dagger + \pmatrix{A & 0 & 0 \cr 0 & 0 & 0 \cr 0 & 0 & 0} \ , 
\label{effec_hamilt}
\eeq
where $A \equiv \pm \sqrt{2}G_F N_e$ for $\barnue$,
with $N_e$ being the instantaneous electron density of the matter.

The survival probablities suffer a
consequent modification. While the full expressions are cumbersome,
it is instructive to look at a
simplified form obtainable for a matter density small enough that $A$
can be treated as a perturbation in the effective Hamiltonian. In this
limit, following Ref. \cite{choubey_roy},
\beq
\barr{rcl}
P_{\mu\mu \, (\bar \mu \bar \mu)}
     & \simeq & \dis P^{\rm vac} \pm (\Delta P_{\mu \mu} / 2)
     + {\cal O}(\atil^2)
  \\[1.5ex]
\Delta P_{\mu\mu} &\simeq&
4 \, \atil \, |U_{e3}|^2 |U_{\mu 3}|^2
(1-2 |U_{\mu 3}|^2)  \,
\left[4  \sin^2 \Psi - \Psi \, \sin (2 \, \Psi) \right]
     + {\cal O}(\atil^2)
\earr
   \label{survival_matter}
\eeq
where
\beq \dis
\atil \equiv  \frac{A}{\Delta_{31}}
\qquad {\rm and} \qquad
\Psi  \equiv  \frac{\Delta_{31} \, L}{2} \ .
\eeq
The asymmetry $\Delta P_{\mu\mu}$ is a measure of the matter effect
and that it should be proportional to both $\atil$ and $|U_{\mu 3}|^2$ is
obvious. Note also that
the $\nu_\mu$ has the same interaction with matter as the $\nu_\tau$
with which it primarily mixes and any matter effect can only seep in
through a mixing with the $\nu_e$. This, then, accounts for the overall
factor of $|U_{e3}|^2$. All of these three proportionalities quite 
independent of the approximation, and, in fact, are exact
results. The  factor $(1-2 |U_{\mu 3}|^2)$,
however, is but a consequence of the approximation of a small
matter term $\atil$ and is applicable only for neutrinos
traversing small base lines. Since this factor vanishes identically
for a maximal mixing in the $\nu_\mu - \nu_\tau$ sector,
this matter-induced asymmetry is potentially  a sensitive probe
of the maximality of $U_{\mu 3}$. This has  already been pointed
out\cite{choubey_roy} and examined in the context of the
{\sc minos} experiment.
In the next section we will briefly reexamine their
claims.

Convoluting eqn.(\ref{survival_matter})
with the $E_\nu$-dependent flux and the detection efficiency of a
given detector, one would obtain the number of events and thereby
the event asymmetry
\beq
\Delta N = N_{\nu_\mu} - N_{\bar \nu_\mu} \equiv N_- - N_+ \ .
\eeq
The latter could then be used to place contraints in the
$|U_{\mu 3}|$--$|U_{e 3}|$ plane.
Of course, in a realistic case,
the effects of a varying matter density has to be taken into account
as well and this we do include in our analysis. With the
effect being proportional to $\atil$, it will be magnified when the
(anti-)neutrino beam traverses regions of high density, or in other words,
the core of the earth. Note, however, that a large $\atil$ results in a
breakdown of the simplified form given in eq.(\ref{survival_matter}) and
that the full expression (or, at least, a different
approximation) needs to be used.

The exact expression (for an arbitrarily large but constant density) for the
survival probability is given by
\beq
\barr{rcl}
P_{\mu\mu} & = & \dis
1 - 4 \, s^2_{23}\, \Bigg( c^2_{23} s^2_{\theta_m}
    \sin^2 \left[ {\Delta_{31} L \over 4} \,
             \left( 1 + \atil - {\cal D} \right) \right]
\\[1ex] && \dis \hspace*{4em} +
s^2_{23} s^2_{\theta_m} c^2_{\theta_m}
      \sin^2 {\Delta_{31} L {\cal D} \over 2}
+ c^2_{23} c^2_{\theta_m} \sin^2
\left[ {\Delta_{31} L \over 4} \,
             \left( 1 + \atil + {\cal D} \right) \right] \Bigg) \ ,
\earr
\eeq
where
\beq \dis
{\cal D} \equiv \sqrt{1 + \atil^2 - 2 \, \atil \cos 2 \theta_{13}}
\ , \quad {\rm and} \quad
\theta_m \equiv  \frac{1}{2} \tan^{-1}\frac{\sin2\theta_{13}}
{\cos2\theta_{13}-\atil} \ .
\eeq

We now perform a double 
expansion\footnote{This is similar in spirit to, though not the same as, 
   the expansion performed in Ref.\protect\cite{expand}.} 
in $\ue$ as well as
$
\beta \equiv \frac{1}{2} - \Um^2
$
while allowing for any value for $\atil$. Since both the expansion
parameters are small, we may retain terms only upto, say, the fourth order
without any loss of accuracy. Then, for the survival probability, we have
\beq
\barr{rcl}
P_{\mu \mu} & = & \dis
\left(1 - 4 \, \beta^2 \right) \, \cP^2
\\[2ex]
& + & \dis
  \frac{ \ue^2 }{{\left( 1 - \atil \right) }^2}
\,\Bigg[
    \Bigg\{ \sP^2 - \sOm^2  -  s_\Theta^2
 - \atil \, \left( 1 - \atil \right) \,
\Psi\,
         s_{2 \Psi}    \Bigg\}
- 4 \beta \;
          \Bigg\{  (1 - \atil)^2 \, \sP^2
- \sOm^2
\Bigg\}
\\[2ex]
&  & \dis \hspace*{4em}
+
     4\, {\beta}^2
\, \Bigg\{ (1 - 4 \atil + 2 \atil^2) \,
   \sP^2
+
        \sOm^2  -  s_\Theta^2
+  \frac{\atil}{2} \left( 1 - \atil \right) \,\Psi\,
           s_{2 \Psi}  \Bigg\}
\Bigg]
\\[2ex]
& + & \dis
  \frac{\ue^4}{
     {\left( 1 - \atil \right) }^4}\,\Bigg[
 \left(1 - 6\,\atil  + {\atil}^4
          -  4\,{\atil}^3 + 5 {\atil}^2\right) \,
                        \sP^2
+
       (2\,\atil + {\atil}^2) \, \sOm^2
\\[2ex]
&  & \dis \hspace*{6em}
 -  \Omega^2 \, (1 - \atil)^2 \, c_{2 \Psi}
 + ( -1 + 6\,\atil - {\atil}^2) \,
 s_\Theta^2
\\[2ex]
&  & \dis \hspace*{6em}
+
       \left( 1 - {\atil}^2  \right) \, \Omega\,  s_{2 \Psi}
+
       \left(1 - {\atil}  \right) \, \Omega\,
             \left( s_{2 \Omega} + 2 \, s_{2 \Theta} \right)
         \Bigg]
\\[2ex]
& + & \dis {\cal O}(\ue^4 \beta, \ue^3 \beta^2, \ue^2 \beta^3, \ue \beta^4) \ ,
\earr
   \label{survival_simp}
\eeq
where $s_\alpha \equiv \sin (\alpha)$,  $c_\alpha \equiv \cos (\alpha)$
with
\beq \dis
\Omega \equiv \frac{\Delta_{31} \, L \, \atil}{2} = \frac{A \, L}{2}
        = \atil \, \Psi
\quad
{\rm and}
\quad
\Theta \equiv \Omega - \Psi \ .
\eeq
With the survival probability for the antineutrinos being
obtained by the replacement $\atil \to - \atil$ in
eq.(\ref{survival_simp}), we have, for the probability asymmetry
\beq
\barr{rcl}
\dis
\frac{{- \left( 1 - {\atil}^2 \right) }^3 \, \Delta P}
{4\, \atil \, \ue^2\, \Um^2}
& = & \dis
 ( 1 - {\atil}^2) \;  \Bigg[4\,
     \cP^2  \,  \sOm^2
+ ( 1 - {\atil}^2) \,
\Psi\,s_{2 \Psi} \,
- \frac{1 + \atil^2}{2 \, \atil}  \,
     s_{2 \Psi} \, s_{2 \Omega} \Bigg]
\\[2ex]
& + & \dis
  2\, \beta\,( 1 - {\atil}^2) \,
\Bigg[
 -4\,   \sP^2 \, \sOm^2
+  ( 1 - {\atil}^2) \,\Psi\, s_{2 \Psi}
+ \frac{ 1 + {\atil}^2}{2 \, \atil} \,s_{2 \Psi} \, s_{2 \Omega}
        \Bigg]
\\[2ex]
& + & \dis
   \ue^2\,\Bigg[ \frac{4}{1 - \atil^2}
                    \,
                 \left\{( 1 + 11\,{\atil}^2 + 2 \, {\atil}^4 ) \,
                       \sP^2 \, \cOm^2
+    \atil^2 \, ( 3 - \atil^2 ) \,   \cP^2 \, \sOm^2
                 \right\}
\\[2ex]
& & \dis \hspace*{3em}
+ \Psi\,
\left\{
       2 \, {\atil}^2 \, (1 - {\atil}^2)^2\,   \Psi \, c_{2 \Psi}
-  (1 + 6 \, \atil^2 + \atil^4) \, \sP \, \cP \right\}
\\[2ex]
& & \dis \hspace*{3em}
+ s_{2 \Psi} \,
   \left\{ 2\,  (1 + 3 \, \atil^2) \,
  \Psi\, c_{2 \Omega}
-  \,
\frac{ 1 - 16 \, \atil^2 - \atil^4}{4\,  \atil} \, s_{2 \Omega}\right\}
\\[2ex]
& & \dis \hspace*{3em}
-  (3 + \atil^2) \, \Omega \, s_{2 \Omega} \,
\left(3 - 4\, \sP^2 \right)
\Bigg]
\\[2ex]
& + & \dis {\cal O}(\ue^2 \beta, \ue \beta^2, \beta^3) \ .
\earr
   \label{asym_simp}
\eeq
Note that
\begin{itemize}
\item The asymmetry $\Delta P$ continues to be proportional to each of 
      $\atil$, $\ue^2$ and $\Um^2$, but not to $(1 - 2 \Um^2)$.
\item $\lim_{\atil \to 0}  \sOm / \atil = \Psi$ and hence the
     right hand side of eq.(\ref{asym_simp}) is finite in this limit.
\item To ${\cal O}(\atil)$, the expressions for both $P_{\mu \mu}$ and
$\Delta P$ coincide with those of eq.(\ref{survival_matter}).
\item The strength of the ${\cal O}(1)$ term on the r.h.s. of
eq.(\ref{asym_simp})---which vanishes as $\atil^2$ for small $\atil$---
gives a measure of the violation of the $(1 - 2 \Um^2)$
proportionality of $\Delta P$.
\item While we have performed the expansion upto ${\cal O}(\ue^4)$, for all
practical purposes it suffices to consider only upto  ${\cal O}(\ue^2)$.
\item The apparent singularities in
   eqs.(\ref{survival_simp} \& \ref{asym_simp})  are not
   real ones with
\[
\barr{rcl}
\dis \lim_{\atil \to 1} P_{\mu \mu} & = & \dis
\cP^2
+ 4 \, {\beta}^2\, \sP^2
\\[2ex]
& + & \dis
 \ue^2\,\Bigg[\Psi\, (s_{2 \Psi} - 2 \, \Psi \, \cP^2)
         + 4 \, \beta\,\left( \Psi^2 - \sP^2 \right)
+    4 \, {\beta}^2\,\left\{ 2 \, \left( 1 - \Psi^2 \right)  \, \sP^2
         - \Psi\, s_{2 \Psi} \right\} \Bigg]
\\[2ex]
& + & \dis
  \ue^4\,\Psi^2 \, \left[ \sP^2 +
     \frac{\Psi^2}{3} \, (3 + 2 \, \sP^2 )
     - \frac{4}{3} \, \Psi\,s_{2 \Psi} \right] \ ,
\earr
\]
and
\[
\barr{rcl}
\dis
\lim_{\atil \to 1} \frac{- \Delta P}{\ue^2\, \Um^2}
& = & \dis
 2 \, \Bigg[ 2 \, \Psi^2\, \cP^2
-      \Psi\,\sP\,\cP -
               \frac{s_{2 \Psi}^2}{4}    \Bigg]
+
\beta \,\Bigg[ -        \Psi\,s_{2 \Psi}
+  (5   - 8 \, \Psi^2) \, \sP^2
+ s_{2 \Psi}^2  \Bigg]
\\[2ex]
& + & \dis
  \ue^2\,\Bigg[  -s_{2 \Psi}^2
           -   \Psi^2\, \left(\frac{3}{2} +  \sP^2 \right) -
        \frac{2 \, \Psi^4}{3} (3 + 2 \, \cOm^2)
 \\[2ex]
 & & \dis \hspace*{4em}
+ \frac{4\,\Psi^3}{3}\, s_{2 \Psi} +
        \Psi\,s_{2 \Psi} \,
        \left( \frac{5}{8} - \sP^2 \right)
  \Bigg] \ .
\earr
\]

\end{itemize}

\section{Can the NuMI--MINOS Combine See This?}
     \label{sec:minos}
With the Fermilab Main Injector serving as a neutrino source,
the {\sc minos} detector~\cite{minos} at the Soudan Mine offers a
baseline of 732 km and hence a possible setup for making such a
measurement. This was investigated by Choubey and Roy in
Ref.\cite{choubey_roy}. Working with a
$\nu_\mu$ exposure of $16\times 10^{20}$ primary protons on target, they
conclude that a
significant rate asymmetry ($\Delta N$) can be observed for moderate values
of $(1 - 2 |U_{\mu 3}|^2)$ and $|U_{e 3}|^2$.

However, while their idea was an interesting one 
and the formalism essentially correct, a reestimation of the rate 
asymmetry using the
same set of inputs  reveals some discrepancies.  While we agree with
Ref.\cite{choubey_roy} on the magnitude of $\Delta N$ for all the
different cases of inputs parameters, we completely disagree on the
issue of the error bars. Even on using an optimistic indicator of the
background fluctuation, namely $\sqrt{N_\mu + N_{\bar \mu}}$, we find
that, for the abovementioned luminosity, the statistical error bars
are large enough for each of the rate asymetries in
Ref.\cite{choubey_roy} to be consistent with zero. 

\begin{figure}[htb]
\vspace*{-20ex}
\centerline{\epsfig{figure=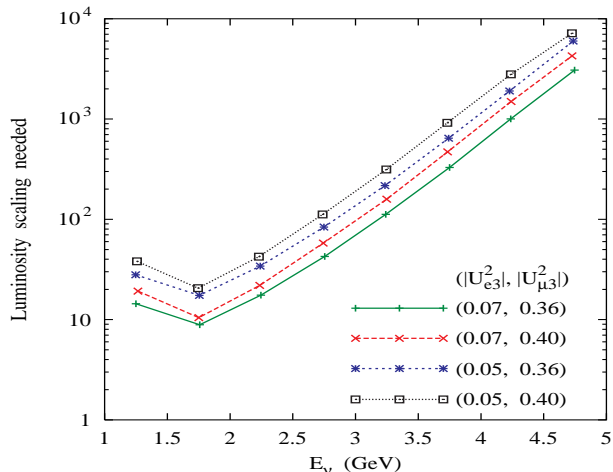,height=4.0in,width=3.5in,angle=0}}
\vspace*{-5ex}
\caption{\em The minimal ratio by which the {\sc minos} $\nu_\mu$ exposure
($16\times 10^{20}$ primary protons on target) needs to be scaled up
to obtain $1\sigma$ effects for each individual energy bin.
The curves correspond to the four parameter points of
Ref.\protect\cite{choubey_roy}. Note that only the statistical
fluctuations have been included in this analysis.}
\label{figure:flux_scaling}
\end{figure}
%

Thus, to obtain a
result of any statistical significance would need the flux (or, equivalently,
the detector size) to be
increased manifold (see Fig.~\ref{figure:flux_scaling}).
Such a large scaling,
unfortunately, is not feasible with the present experimental facility.
Note, in addition, that
this projection altogether neglects the systematic uncertainties, and
consequently is already too optimistic.
In reality, several uncertainties are involved 
in the translation between the visible energy observed at a detector 
and the energy of the neutrino to which the event is ascribed. These 
include the uncertainties in the assumed final state multiplicities, the 
scattering or absorption of the secondary particles etc. In a 
recent paper~\cite{minerva},
the MINER$\nu$A collaboration, estimates the size of the corresponding 
systematic  effects at {\sc minos}-like experiments could be
comparable to, or even dominate, the statistical errors.

\section{The case for a very long baseline}
      \label{sec:longbaseline}

Having established that, contrary to the claims of Ref.\cite{choubey_roy},
{\sc minos}, in its current incarnation,
is not sensitive to possible deviations of $\Um$
from maximality even for optimistic choices of parameters, it becomes
interesting to investigate if any of the other proposed long-baseline
experiments would do the job. And, in a similar vein, if even the
luminosity scaling as described in the previous section would prove adequate.
We start by investigating the second problem.

\begin{figure}[hbt]
\vspace*{-0.4cm}
\centerline{
\hspace*{3em}
\epsfxsize = 10cm  \epsfysize = 7cm \epsfbox{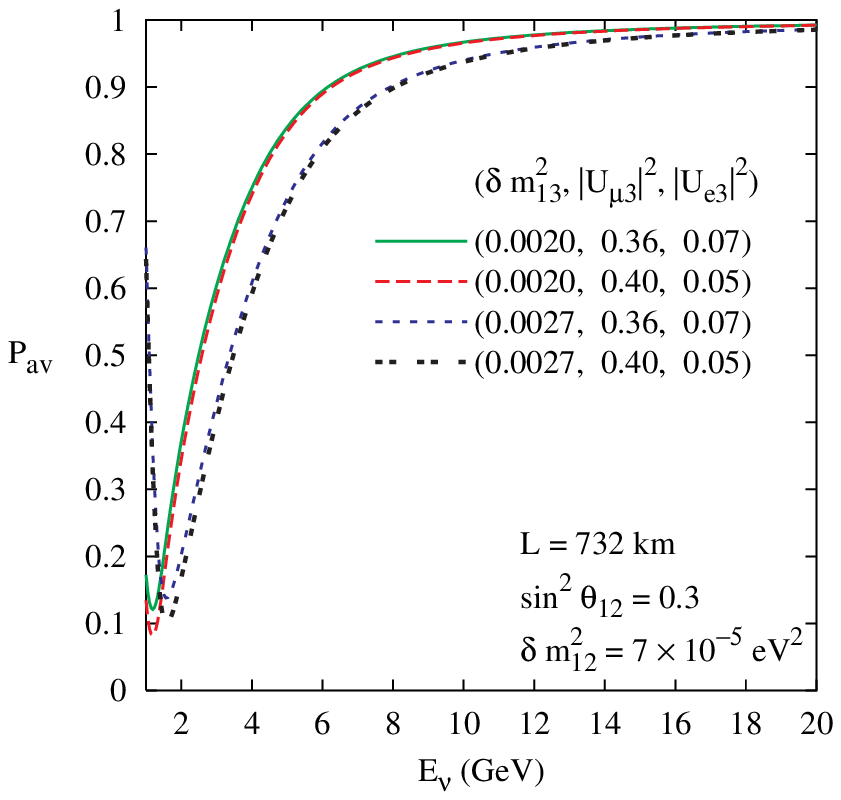}
\hspace*{-7em}
\epsfxsize = 10cm  \epsfysize = 7cm \epsfbox{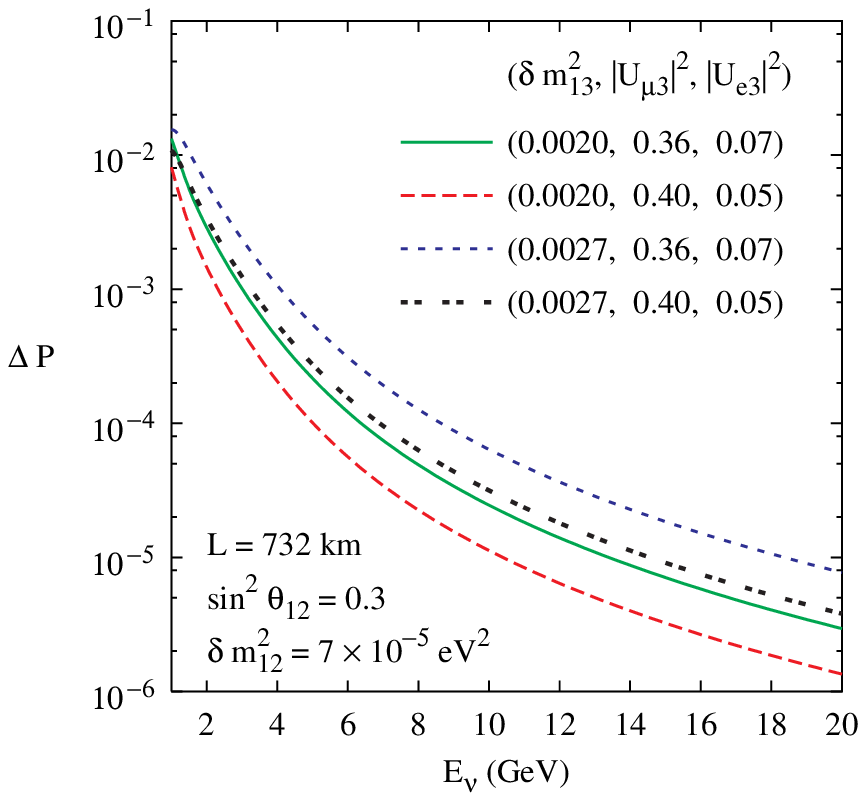}
}

\vspace*{-0.8cm}
\caption{\em {\em (a)} The neutrino survival probability
as a function of neutrino energy for a baseline $L = 732$ km.
The curves correspond to different combinations of
$\delta m^2_{13}$ (in $\ev^2$), $|U_{\mu 3}|^2$ and $|U_{e 3}|^2$.
{\em (b)} The corresponding difference in the $\nu_\mu$ and
$\bar\nu_\mu$ survival probablities.}
  \label{Fig:prob_732}
\end{figure}

As has already been emphasised above, 
it is not only $\Delta P$ that is important,
but also the average survival probability
$P_{\rm av} \equiv (P_{\mu \mu} + P_{\bar\mu \bar\mu}) / 2$ for the latter
controls the total number of events and hence the size of the statistical
error (significance
$\propto \Delta P_{\mu \mu} / \sqrt{2 P_{\rm av}}$). In
Fig.\ref{Fig:prob_732}, we present both these quantities as function of
neutrino energy for a baseline of 732 km. The two choices of
$(U_{\mu 3}|^2, |U_{e 3}|^2)$ correspond to the most (and least) optimistic
sets of parameters used in Ref.\cite{choubey_roy}.

It is abundantly clear that, for such a baseline, and for
realistic choice of parameters,
$\Delta P$ is very small, especially at larger neutrino energies. This
was to be expected as a baseline of 732 km implies that the neutrino
beam neither travels through a very dense matter core nor does it
travel through a ``dilute'' segment long enough for the matter effect
to build up sufficiently. Clearly, it is dangerous to claim
a signal based on $\Delta P \lsim 10^{-4}$, for such an act presupposes
a very accurate knowledge of both the (anti-)neutrino cross sections
as well as the density profile within the earth. Thus, one would have to
concentrate only on neutrinos with relatively small energies
($E_\nu \lsim 2 \gev$). Apart from the fact that  the
asymmetry is not too large even in this 
range\footnote{In fact, it stands to be swamped by the systematic 
   errors unless further experiments are performed~\protect\cite{minerva}.},
such a measurement also necessitates both a very good energy
resolution as well as a low energy threshold.
Furthermore, for such small energies, neutrino cross-section
scales as $E_\nu^2$ and thus restricting ourselves to a small
window is tantamount to rejecting a very large fraction of the events.
\begin{figure}[hbt]
\vspace*{-0.3cm}
\centerline{
\hspace*{3em}
\epsfxsize = 10cm  \epsfysize = 7cm \epsfbox{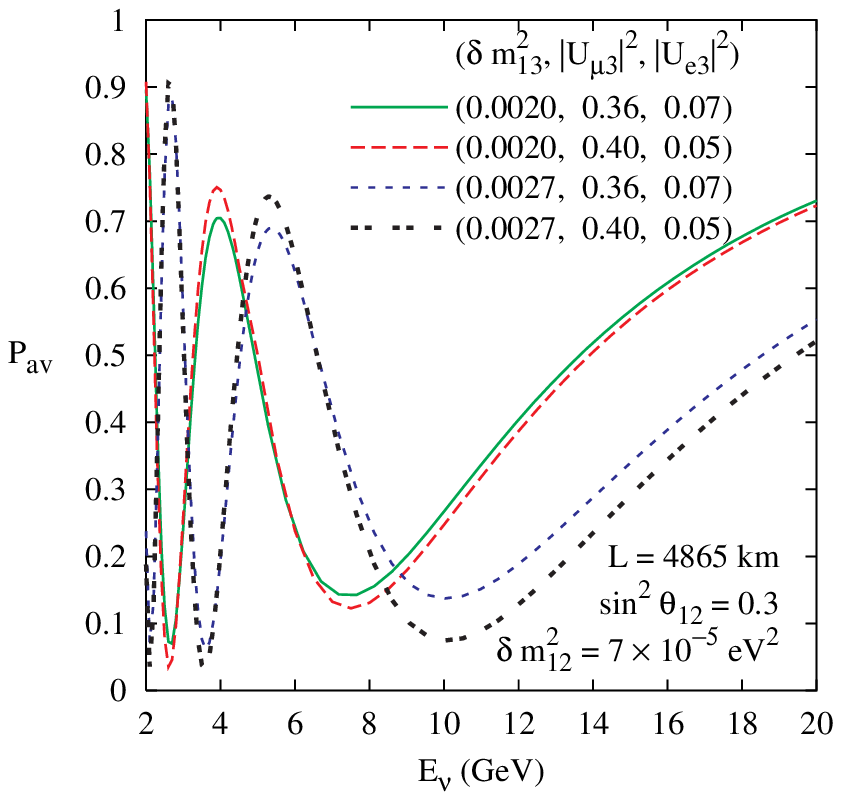}
\hspace*{-7em}
\epsfxsize = 10cm  \epsfysize = 7cm \epsfbox{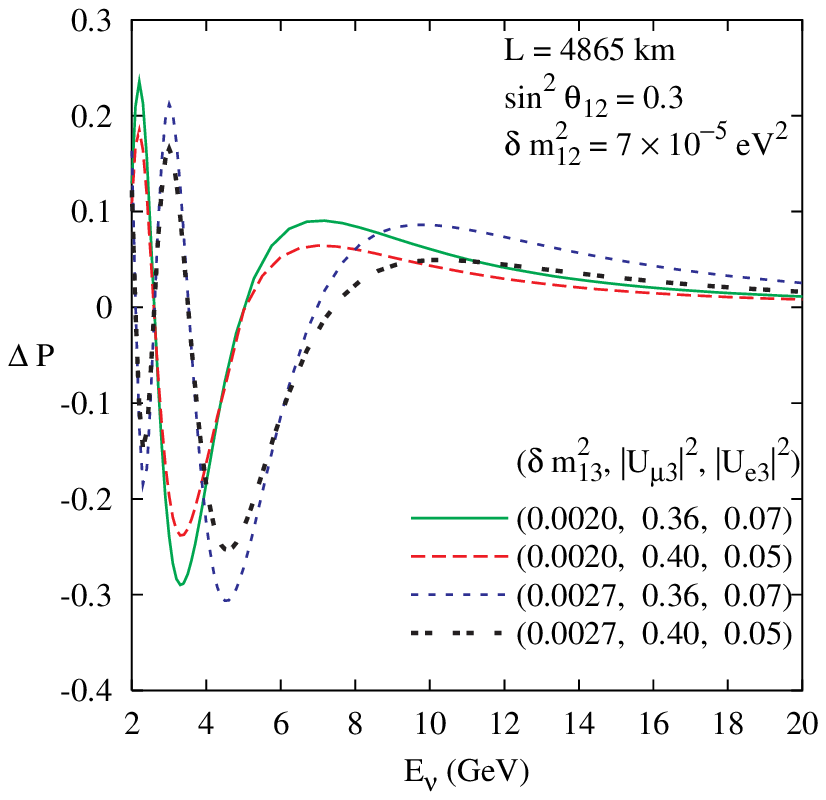}
}

\centerline{
\hspace*{3em}
\epsfxsize = 10cm  \epsfysize = 7cm \epsfbox{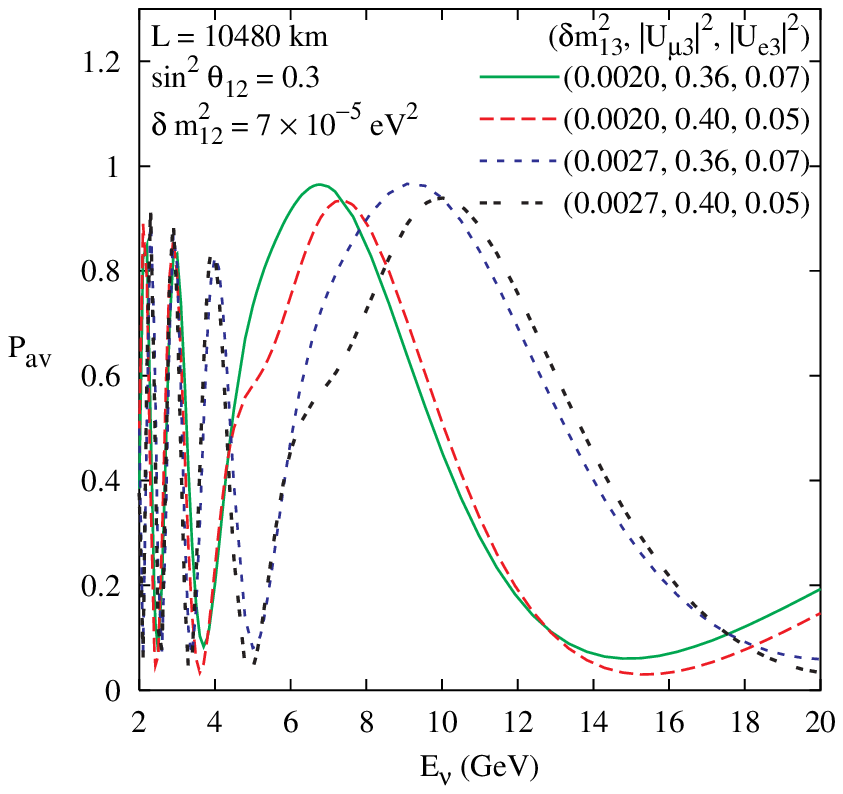}
\hspace*{-7em}
\epsfxsize = 10cm  \epsfysize = 7cm \epsfbox{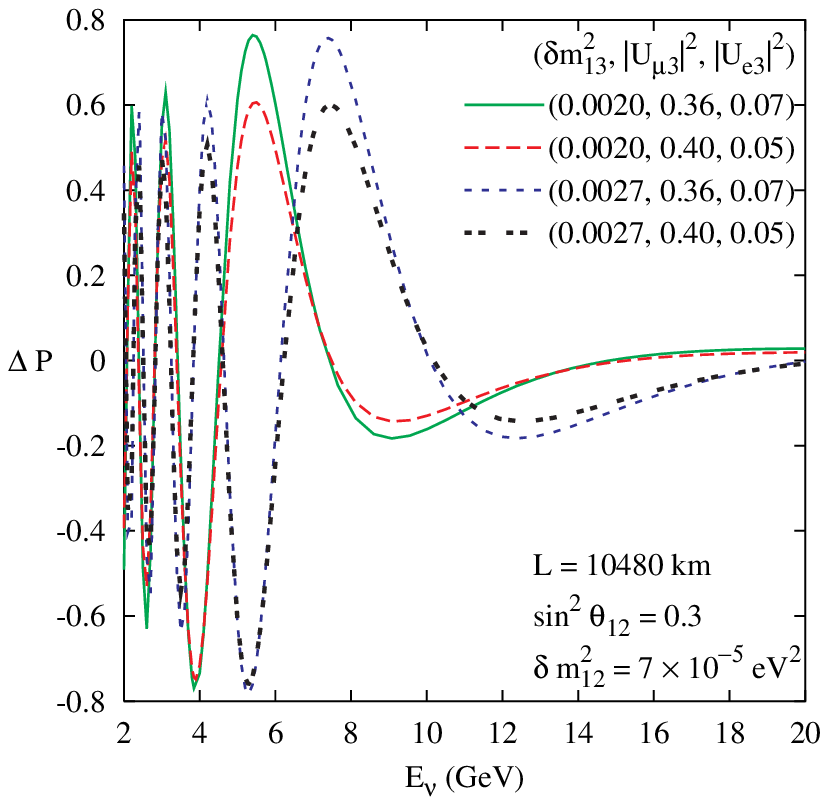}
}

\vspace*{-0.8cm}
\caption{\em As in Fig.\protect\ref{Fig:prob_732}, but for baselines
$L = 4865$ km (upper panels) and  $L = 10480$ km (lower panels) instead.
}
  \label{Fig:prob_4865_10480}
\end{figure}

The situation, vis. a vis. the size of $\Delta P$, seems to improve
dramatically with an increase in the baseline.
For definiteness, we consider two particular values, namely
$L = 4865$ km (the distance between the Japan Hadron Facility and the
proposed Indian Neutrino Observatory site at Rammam~\cite{ino}) and
$L = 10480$ km (the distance between Fermilab and INO).
Since the matter term can no longer be treated as a perturbation,
 we desist from using the simplified form of
  eq.(\protect\ref{survival_matter}) and use the full expression for neutrino
  propagation in matter with the density
profile being given by the Preliminary Reference Earth Model\cite{prem}.
As Fig.\ref{Fig:prob_4865_10480} amply demonstrates, for each of these
values, $\Delta P$ is large over a wide range of $E_\nu$. In fact, for
the larger of the two baselines $|\Delta P|$ can be as large as 0.8 for
certain $E_\nu$ bins. Furthermore, unlike in the case of
the 732 km baseline where $\Delta P$ was nearly monotonic in $E_\nu$,
significant structure, including sign reversal, is shown. As can be easily
appreciated, the latter feature has the potential of serving as a key
experimental signature.

A further point of interest relates to the position of the maxima
of the survival probabilities as well as the asymmetry. Clearly, a
high resolution measurement of the same could, in principle, be
used to determine $\delta m_{13}^2$.

\begin{figure}[hbt]
\vspace*{-0.3cm}
\centerline{
\hspace*{3em}
\epsfxsize = 10cm  \epsfysize = 7cm \epsfbox{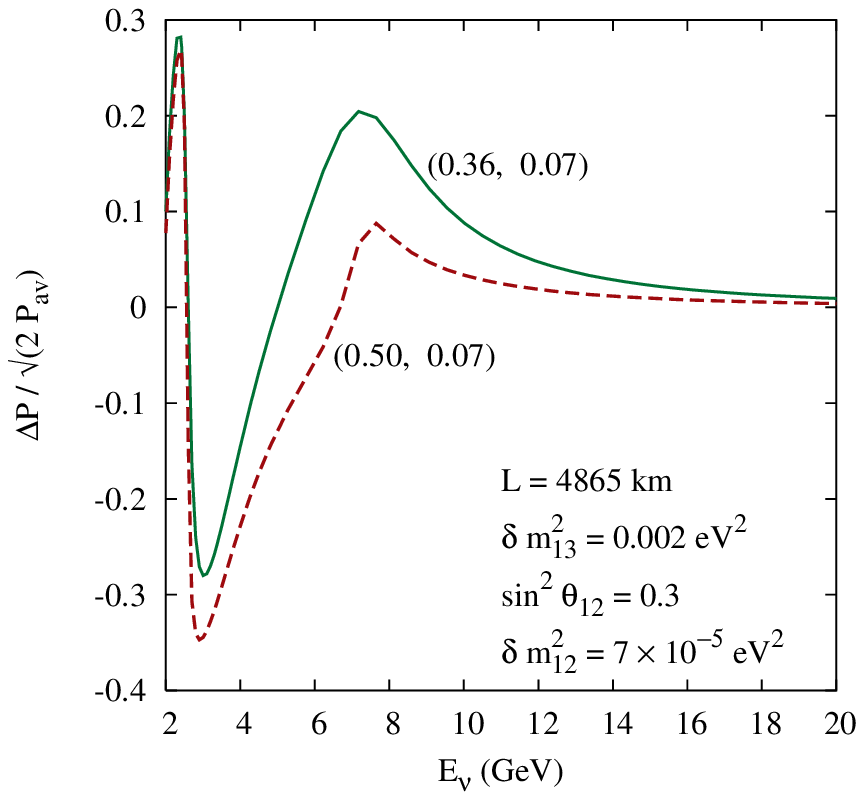}
\hspace*{-7em}
\epsfxsize = 10cm  \epsfysize = 7cm \epsfbox{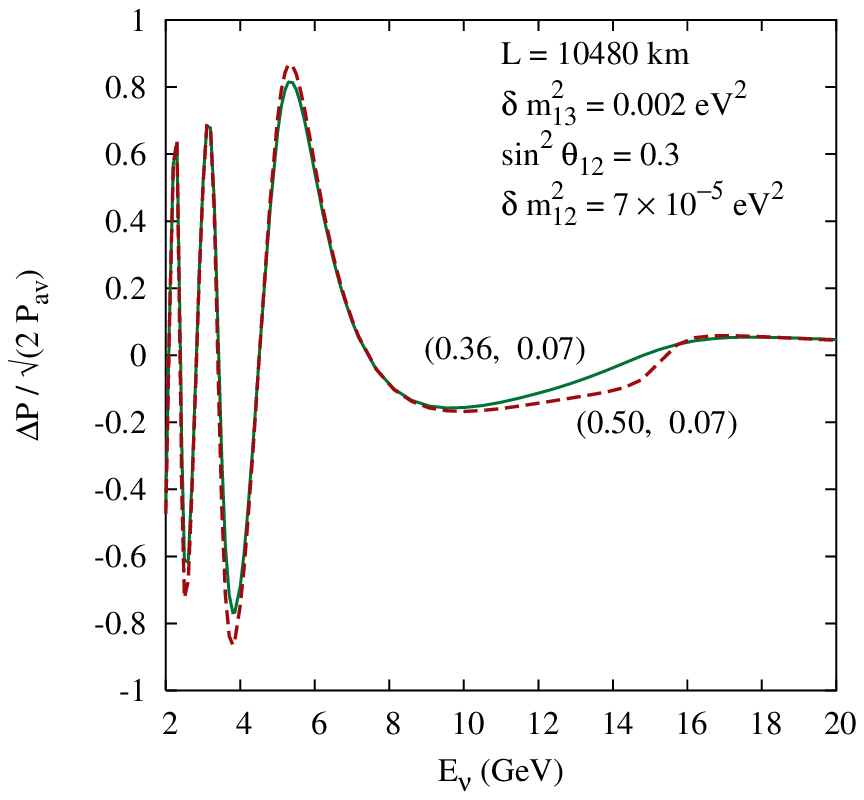}
}
\vspace*{-0.8cm}
\caption{\em The normalized probability asymmetry for a baseline of
 {\em (a)} 4865 km and {\em (b)} 10480 km. In each case, the solid
 (dashed) line refers to $|U_{\mu 3}|^2 = 0.36$ ($0.50$).
 Note that, according to the leading expression of
 eq.(\protect\ref{survival_matter}), the dashed curves should have
 coincided with the abscissa.
}
  \label{Fig:comp_4865_10480}
\end{figure}

Before we interpret Fig.~\ref{Fig:prob_4865_10480} to imply that
the asymmetry measurement in very long baseline experiments would
be a sensitive probe of the deviation of $|U_{\mu 3}|$ from maximality,
it should be realised that the latter is not the only source
for a non-zero asymmetry. In fact, as has already been hinted at
above, the higher order terms in $\atil$ that were dropped
while deriving eq.(\ref{survival_matter}) are {\em not} proportional to
$(1 - 2 |U_{\mu 3}|^2)$. This can also be seen explicitly from
eqs.(\ref{survival_simp} \& \ref{asym_simp}) obtained
assuming a constant density profile for earth.
The numerical importance of such contributions
naturally increases with the baseline,
 and has been explicitly displayed in
Fig.~\ref{Fig:comp_4865_10480}. In fact, for the larger baseline, it is
quite apparent that the bulk of the effect is due to such ``higher-order''
terms. Thus, it may be concluded
that while a baseline of $\approx 5000$ km may allow for a determination of
$(1 - 2 |U_{\mu 3}|^2)$ through measurements of the rate asymmetry,
for baselines much longer than this the sensitivity reduces quite sharply,
and a measurement of the maximality of $\Um$ is not very
straightforward for ultra-long baselines.

On the other hand, the asymmetry must remain
proportional to $\ue^2$ even on the inclusion of $\atil^n$ terms, 
for it is
only through the interaction with the $\nu_e$ that matter effect truly
enters $\nu_\mu$--$\nu_\tau$ oscillations. Thus, a baseline of
10480 km may still serve the purpose of affording a good measurement of
$|U_{e 3}|$. In fact, since $U_{\mu 3}$ and $U_{e 3}$ are inextricably
linked even for the case of the 4865 km baseline, such an independent
measurement of the latter element has its own importance.

\subsection{$\nu_e \to \nu_\mu$ oscillations}

With the matter effect playing such an important role in the survival
probability for the muon (anti-)neutrino, it becomes interesting to
consider its effect on $P(\nu_e \to \nu_\mu)$. For, a source that
produces $\nu_\mu$ ($\bar\nu_\mu$) copiously, whether it be a
beam dump or a muon storage ring, would also produce electron
(anti-)neutrinos.

\begin{figure}[hbt]
\vspace*{-0.3cm}
\centerline{
\hspace*{3em}
\epsfxsize = 10cm  \epsfysize = 7cm \epsfbox{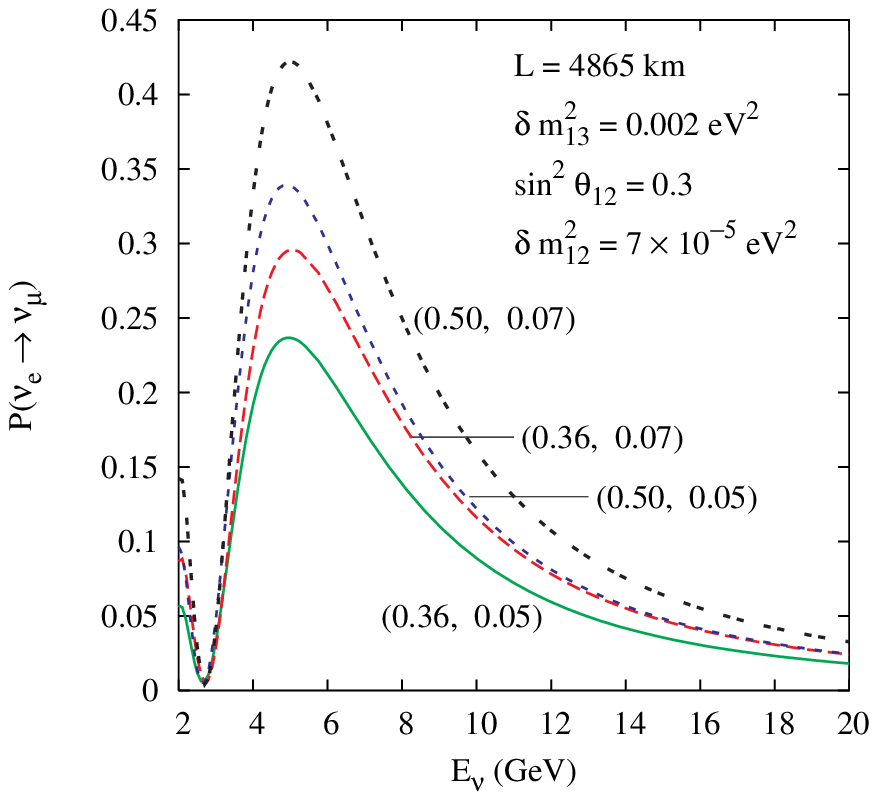}
\hspace*{-7em}
\epsfxsize = 10cm  \epsfysize = 7cm \epsfbox{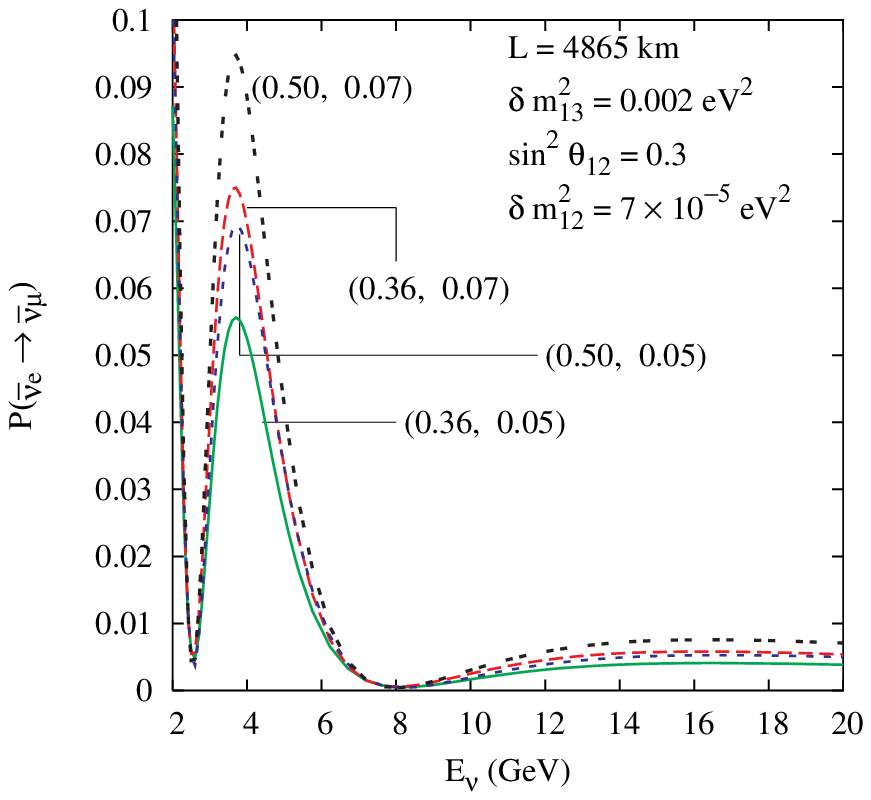}
}

\centerline{
\hspace*{3em}
\epsfxsize = 10cm  \epsfysize = 7cm \epsfbox{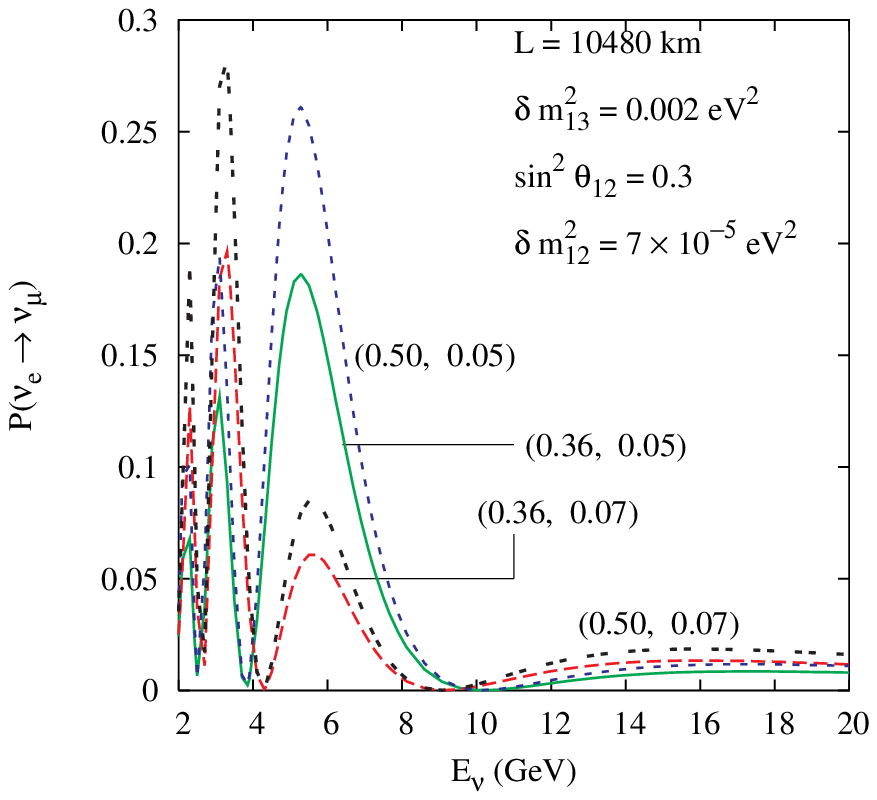}
\hspace*{-7em}
\epsfxsize = 10cm  \epsfysize = 7cm \epsfbox{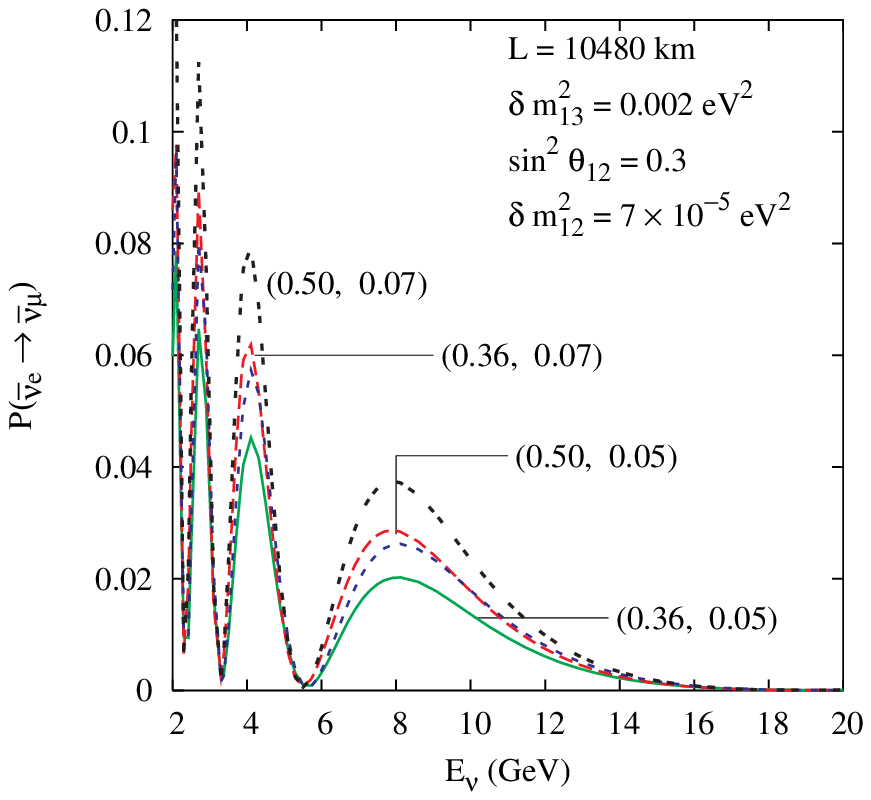}
}
\vspace*{-0.8cm}
\caption{\em The energy dependence of
the transition probability for $\nu_e \to \nu_\mu$ (left)
  and $\bar\nu_e \to \bar\nu_\mu$ (right) on passage through
  matter. The upper and lower panels correspond to baselines
  of 4865 and 10480 km respectively.
}
  \label{Fig:emu_prob}
\end{figure}

In Fig.~\ref{Fig:emu_prob}, we present the probabilities for such transitions
as a function of (anti-)neutrino energy. Several observations are in
order:
\begin{itemize}
\item The transition probability $P(\nu_e \to \nu_\mu) $ can be quite sizable
  for parameter values that are still allowed by experimental data.
\item For $E \gsim 3 \gev$, we see that
  $P(\bar\nu_e \to \bar\nu_\mu) \lsim P(\nu_e \to \nu_\mu)$ (except,
  of course, near the node at $\sim 8 \gev$ for the 10480 km case). Coupled
  with the smaller detection efficiency for anti-neutrinos, this indicates
  that it would be more profitable to work with $\nu_e$ than $\bar \nu_e$.
\item The transition probabilities have non-trivial dependence on both
  $U_{e 3}$ and $U_{\mu 3}$. Unlike in the case of the aforementioned
  asymmetry, these are {\em not} in general proportional to $|U_{e 3}|^2$.
\item The measurement of this effect would thus lead to a constraint in the
  $U_{e 3}$--$U_{\mu 3}$ plane independent of that drawn from the asymmetry
  measurement.
\end{itemize}

\section{Event rate calculation  from a storage ring neutrino source}
    \label{sec:storage_ring}
As we have already mentioned in the introductory section, the
neutrino flux from a muon storage ring can be calculated very precisely.
Starting with a $\mu^-$ beam, the number of $\mu ^-$ events in a
far detector can be obtained by folding this flux with the
survival probability $P_{\mu\mu} (L,E)$ and the charged current
cross-section:
\begin{equation}
N_{\mu } = N_{n} \int \sigma(\nu_\mu + N
\rightarrow \mu^-  + X)\; \frac{dN_{\nu}}{dE_{\nu_\mu}} \;P_{\mu\mu} (L,E)
\; dE_{\nu_\mu}
\label{no_of_evts}
\end{equation}
where $N_n$ is the total number of nucleons
present in the fixed target. An analogous expression obtains for
$\mu^+$ events as well. Note though that the $\bar \nu_e$ from
muon decay could also oscillate into $\bar \nu_\mu$ while traversing
through the earth and result in muonic events. However,
they result in {\em wrong sign} muons and thus
can be easily distinguished in a magnetized detector. We shall revert
back to them at a later stage.

To be specific, we shall consider a storage ring with 20 GeV
muons and a 50 kT Iron calorimeter detector one such as the
proposed {\sc monolith}~\cite{mono}
or the {\sc ical/ino} experiment~\cite{ino}. The projected
energy threshold for muon detection is about 2~GeV and the resolution
is expected to be better than 0.5 GeV over the entire range.

Before delving into the actual event profile, several points are in order.
\begin{itemize}
\item As has already been demonstrated in the previous section, the difference
in the survival probalities ($\Delta P_{\mu\mu}$) typically grows with
the baseline. Thus the experimental sensitivity could be expected
to increase with the baseline.
(This is modulo the fact that the sensitivity of the asymmetry
to $(1 - 2 \, \Um^2)$ reduces
for very large baselines, while the sensitivity to $\ue^2$
is retained.)

\item On the other hand, for a fixed detector size,
the effective solid
angle subtended by the detector to the source decreases with
increasing baseline. 
 Consequently the neutrino flux goes down
quadratically with the baseline, thereby resulting in a smaller number
of events and hence larger statistical errors.
This, for example, can be read off from
Fig.~\ref{Fig:noosc}, wherein we have displayed the number of events
expected if the survival probability were to be unity.
\begin{figure}[hbt]
\vspace*{-0.3cm}
\centerline{
\hspace*{3em}
\epsfxsize = 10cm  \epsfysize = 7cm \epsfbox{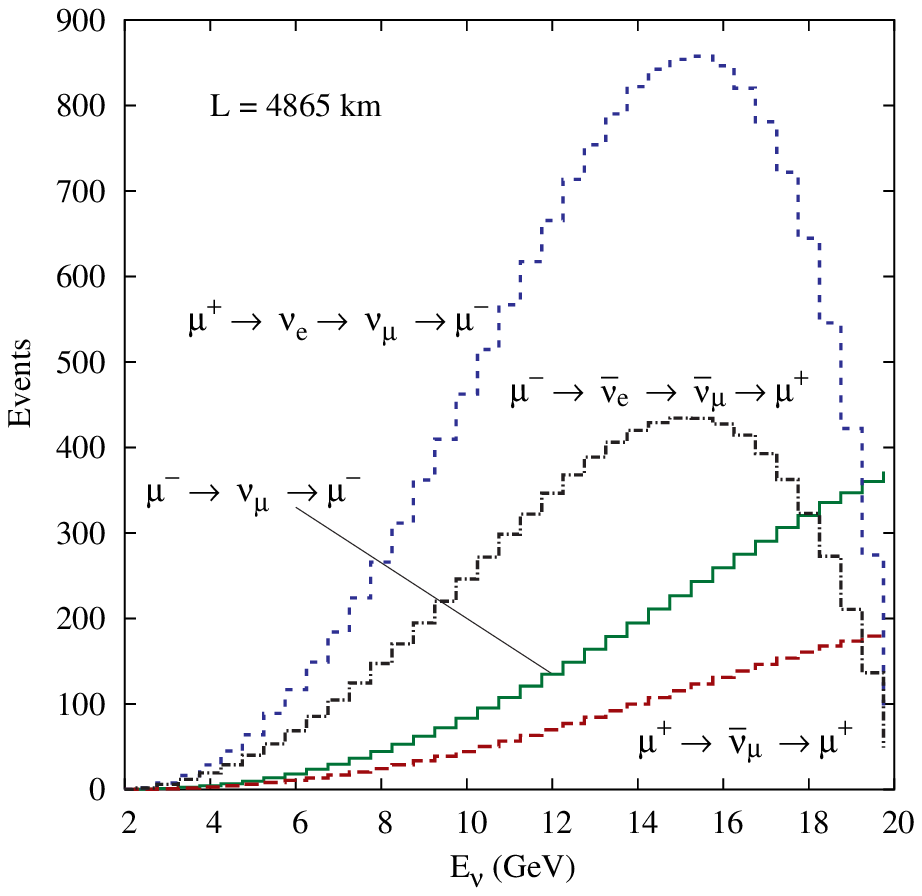}
\hspace*{-5em}
\epsfxsize = 10cm  \epsfysize = 7cm \epsfbox{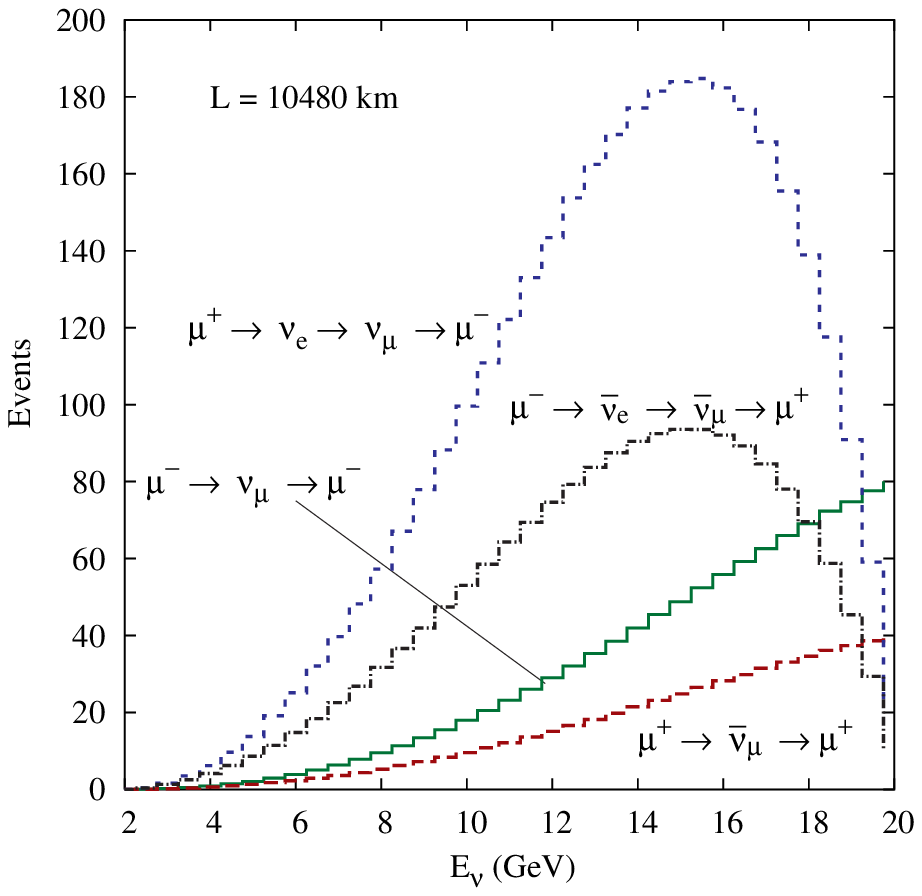}
}

\vspace*{-0.0cm}
\caption{\em The number of events expected as a function of the
(anti-)neutrino energy if the survival probability were to be unity.
The source is a 20 GeV muon storage ring with $10^{20}$ (anti-)muons
decaying while the detector is a 50 kT iron calorimeter with
a energy threshold of 2 GeV for the $\mu^\pm$~\protect\cite{ino}.
The left (right) panel corresponds to a baseline of 4865 (10480) km.
Also shown are the events generated by the electron (anti-)neutrino
assuming a transition probability of unity.
}
  \label{Fig:noosc}
\end{figure}

\item One obvious way to circumvent the latter problem is to start with higher
  luminosity. For our numerical results we then adopt the
  `higher luminosity' option for the muon beam, namely
  $10^{20}$ muon decays per year.

\item An alternative way to the same end would have been to start
   with a higher energy for the muon beam. For example, the neutrino
   beam from a 50 GeV storage ring is collimated sufficiently enough
   to get a similar significance with even the `low luminosity'
   option for the storage ring ($10^{19}$ muon decays per year). However,
   we shall desist from using this option as such a machine is likely
   only as a second generation facility.

\item As Fig. \ref{Fig:prob_4865_10480} amply demonstrates, for longer
   baselines such as the one we are considering, the survival probabilities
   are very sensitive functions of the (anti-)neutrino energy, especially
   for low energies. Thus, extracting any information from the low energy
   tail of the detector muon spectrum would necessitate very good energy
   resolution. However, with the neutrino-nucleon cross sections being
   small for such $E_\nu$, these neutrinos have only a relatively
   small contribution to make to the total number of muon events.
\end{itemize}

Keeping in mind that $\nu -N$ cross-section is nearly
double\footnote{For the total cross section, this ratio has
  only a mild dependence on the incident neutrino
  energy, and for the energy range we are interested in has a value
  close to 2.18. However, on imposition of an
  energy threshold $E(\mu^\pm) > 2 \gev$
  that such a detector is expected to have, the ratio is a more
  sensitive function of the (anti-)neutrino energy
  (see Fig.~\protect\ref{Fig:noosc}).
  In our numerical analysis, we  explicitly account for this
  factor and the consequent effect on the error bars.}
that of $\bar \nu -N$ we argue that the detector should be exposed to
a $\bar\nu_\mu$ beam for double the time that it is exposed to a
$\nu_\mu$ beam. To be specific, we consider a 50 kT-year exposure
to a $\nu_\mu$ beam and a 100 kT-year exposure to $\bar \nu_\mu$ beam,
or in other words, a total exposure of 3 years for the detector
configuration~\cite{ino} under consideration.

\subsection{The rate asymmetry for $\nu_\mu (\bar \nu_\mu)$ initiated events}

\begin{figure}[htb]
\vspace*{-0.2cm}
\centerline{
\hspace*{3em}
\epsfxsize = 10cm  \epsfysize = 7cm \epsfbox{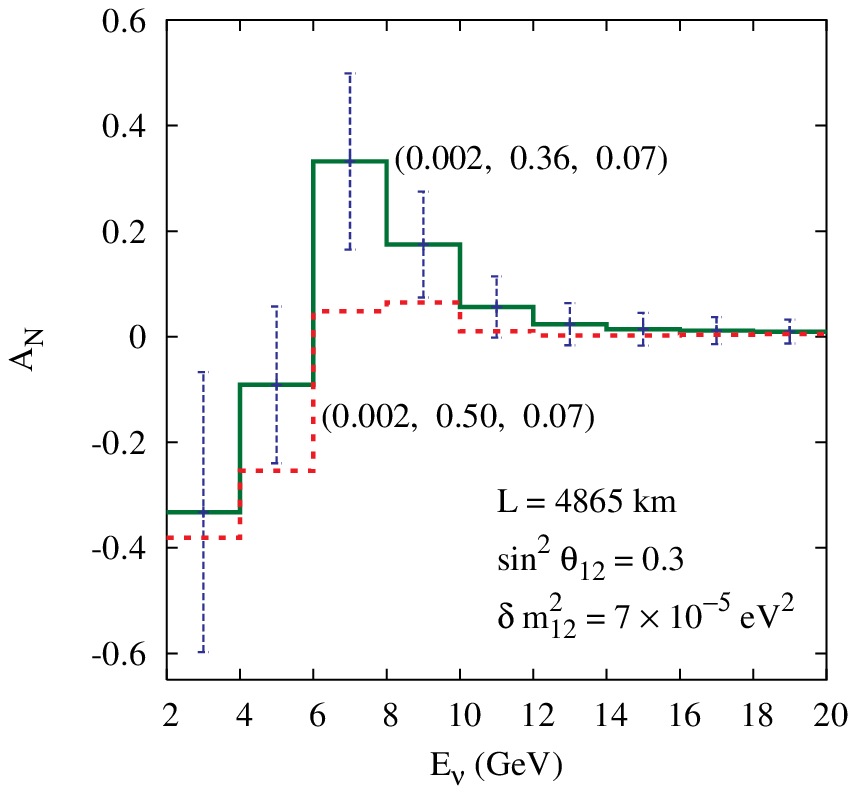}
\hspace*{-7em}
\epsfxsize = 10cm  \epsfysize = 7cm \epsfbox{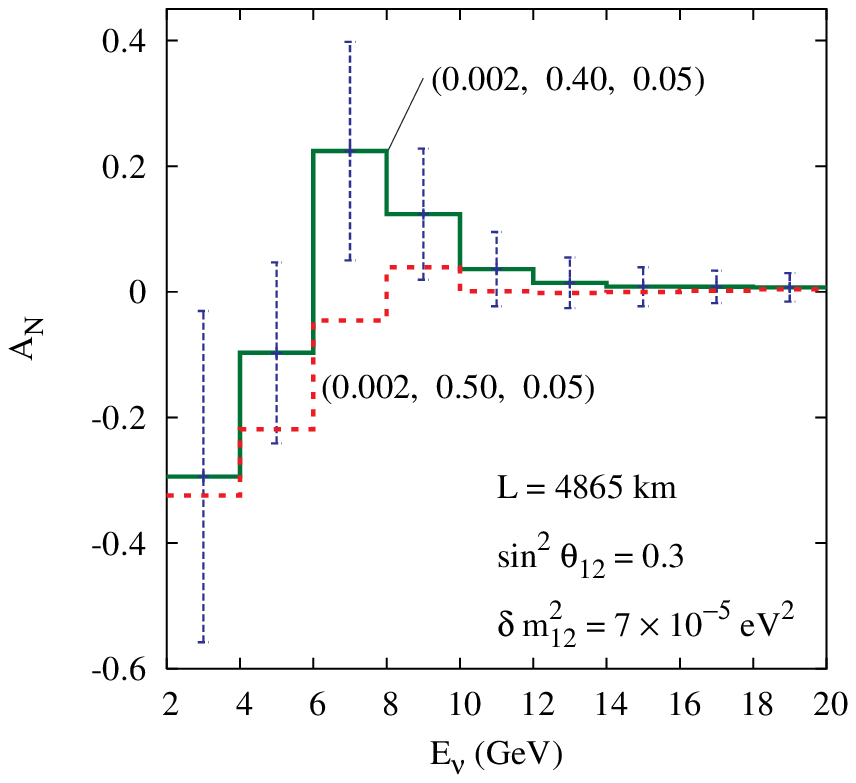}
}

\vspace*{-0.4cm}
\centerline{
\hspace*{3em}
\epsfxsize = 10cm  \epsfysize = 7cm \epsfbox{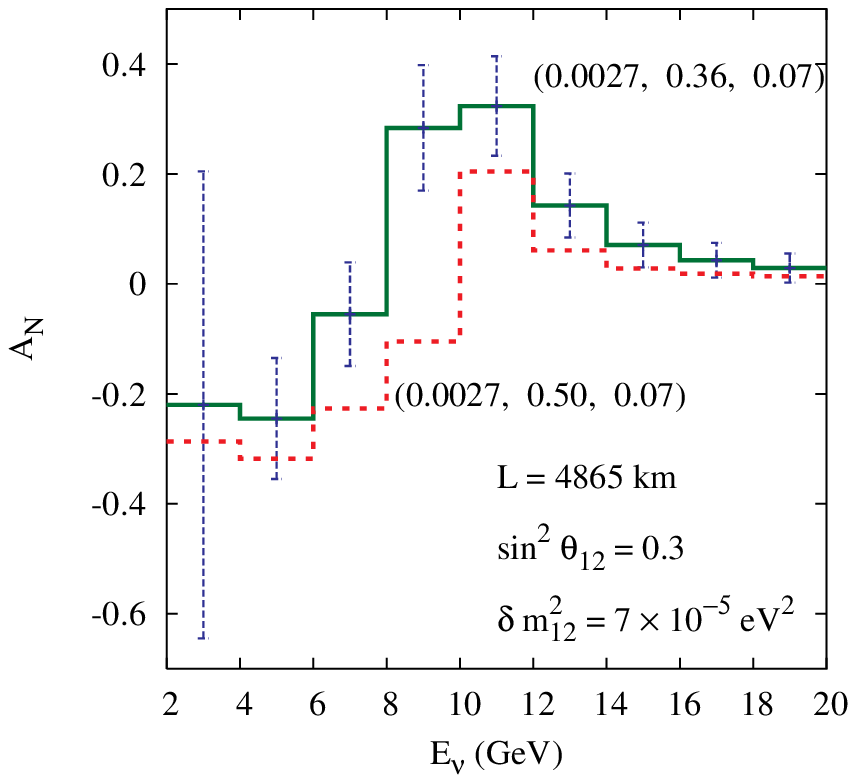}
\hspace*{-7em}
\epsfxsize = 10cm  \epsfysize = 7cm \epsfbox{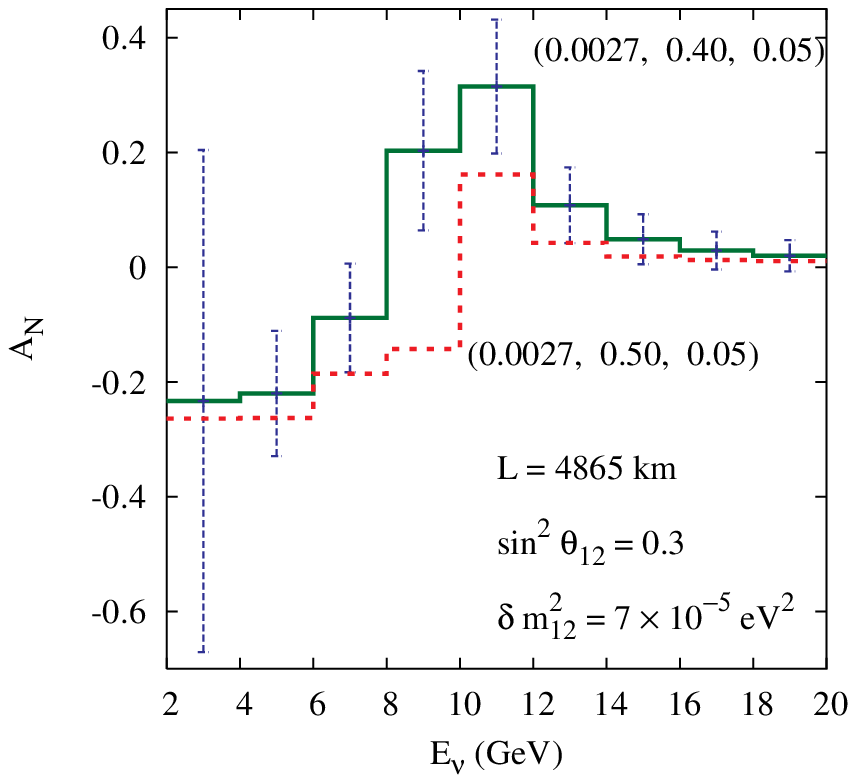}
}

\vspace*{-0.8cm}
\caption{\em The difference between $\nu_\mu$  and $\bar \nu_\mu$
 charged-current events
for a baseline $L = 4865$ km with an integrated luminosity of $10^{20}$ muons
and double the number of antimuons. The detector is a 50 kT iron calorimeter 
with a muon energy threshold of $2 \gev$~\protect\cite{ino}. 
Each plot corresponds to a different
combination of ($\delta m^2_{13}, |U_{\mu 3}|^2, |U_{e 3}|^2$).
The other parameters are:
$\delta m^2_{12} = 7 \times 10^{-5} \ev^2$ and $\sin^2 \theta_{12} = 0.3$.
Also shown are the $1\sigma$ error bars.}
  \label{Fig:diff_4865}
\end{figure}

Armed with the above, we can now calculate the difference in
$\mu^-$ and $\mu^+$ event numbers. Defining an asymmetry
of the form
\beq
A_N \equiv \frac{N_- - N_+}{N_- + N_+} \ ,
   \label{rate_asym}
\eeq
we present, in Fig.~\ref{Fig:diff_4865},  the
same for a baseline of 4865 km. The values for ($|U_{\mu 3}|^2, |U_{e 3}|^2$)
correspond to the two extreme cases of Ref.\cite{choubey_roy}. Also plotted
are the analogous expectations for $|U_{\mu 3}|^2 = 0.5$, a case that, to
the lowest order, would have been associated with vanishing $A_N$.
We have binned the expected data,
keeping under consideration both the expected energy
resolution~\cite{ino} as well
as the number of events in a particular bin. Also shown are the
statistical $1 \sigma$ error bars. Several features are easily
discernible:
\begin{itemize}
\item In each of the cases displayed, the rate asymmetry is statistically
      significant in more than one bin. Together, they lead to a very
      considerable shift in the $\chi^2$. This amply demonstrates the
      sensitivity of such an experiment to a departure from maximality
      of $U_{\mu 3}$.
\item As expected, the signal is less pronounced for the
      case ($|U_{\mu 3}|^2, |U_{e 3}|^2) = (0.40, 0.05)$ than
      for $(0.36, 0.07)$. This is just a reflection of the fact---see
      eq.(\ref{survival_matter})---that the rate asymmetry is proportional to
      $ |U_{e 3}|^2$.  While the proportionality to
      $( 1 - 2 \, |U_{\mu 3}|^2 )$ is no longer an accurate statement, it,
      nevertheless still encapsulates a large measure of truth.

\item In the energy range of interest, the binwise asymmetry clearly
      shows an oscillatory behaviour, a feature that would have been absent
      for a similar detector had the baseline been shorter than $\sim 1000$ km.

\item The functional dependence of $\Delta N(E_\nu)$ exhibits a
      discernible dependence on the mixing angles
      ($|U_{\mu 3}|^2, |U_{e 3}|^2$), and could, in principle, be used
      to distinguish between values for these parameters. However, to
      obtain a significant resolution in the parameter space, a much larger
      event count would be necessary.

\item The dependence on the value of
  $\delta m^2_{13}$\ , on the other hand,  is much more pronounced. The
  shift in $\Delta P$ that we encountered in Fig.\ref{Fig:prob_4865_10480}
  is quite well reflected by a shift in $A_N$, even after the convolution with
  the muon spectrum and the energy-dependent cross sections as well as
  the finite resolution effects. Thus, our previous comment about using
  this measurement for a determination of $\delta m^2_{13}$ is substantiated.

\item The energy dependence is quite different for the two baselines
  considered and can be used to advantage. For example,
  if storage rings come up at both the JHF~\cite{jhf}
  and Fermilab, a detector such as {\sc ical/ino}~\cite{ino} could
  use beams from both to distinguish more efficiently between the
  possible parameter sets.
\end{itemize}

\subsection{Wrong sign muons}

We now turn to a discussion $\nu_e$ to $\nu_\mu$ conversion (as well
the conjugate process) and their detection in the iron calorimeter
detector under consideration. Such events are obviously characterised
by the appearance of a muon with a charge opposite to that of the
decaying particle in the storage ring. Since the detector is to be
magnetized one, charge measurement is relatively straightforward thus
rendering these events easily distinguishable. Efficacy of `wrong-sign-muon'
events in measuring the angle $\theta_{13}$ has already been advocated in
Ref.\cite{barger_matter} in the context of muon-storage ring neutrino factory
 experiments. Here we want to emphasise that how an independent
 measurement of `wrong-sign-muon' events can also supplement the
 more accurate measuremeant of $\theta_{23}$.
We have already seen, vide Fig.\ref{Fig:emu_prob}, that
 the conversion probability for $\nu_e$ is typically larger than that
 for $\bar\nu_e$, a consequence of the sign of $\delta m_{12}^2$. This
 fact, alongwith the smaller detection efficiency for $\bar \nu_\mu$,
 implies that it is wiser to concentrate on the $\mu^+ \to \nu_e \to
 \nu_\mu$ chain rather than attempt to measure the conjugate process
 as well. And since we have already advocated twice as long an
 exposure to $\mu^+$ decays as compared to $\mu^-$ decays, this works
 in our favour too.

\begin{figure}[htb]
\vspace*{-0.2cm}
\centerline{
\hspace*{3em}
\epsfxsize = 10cm  \epsfysize = 7cm \epsfbox{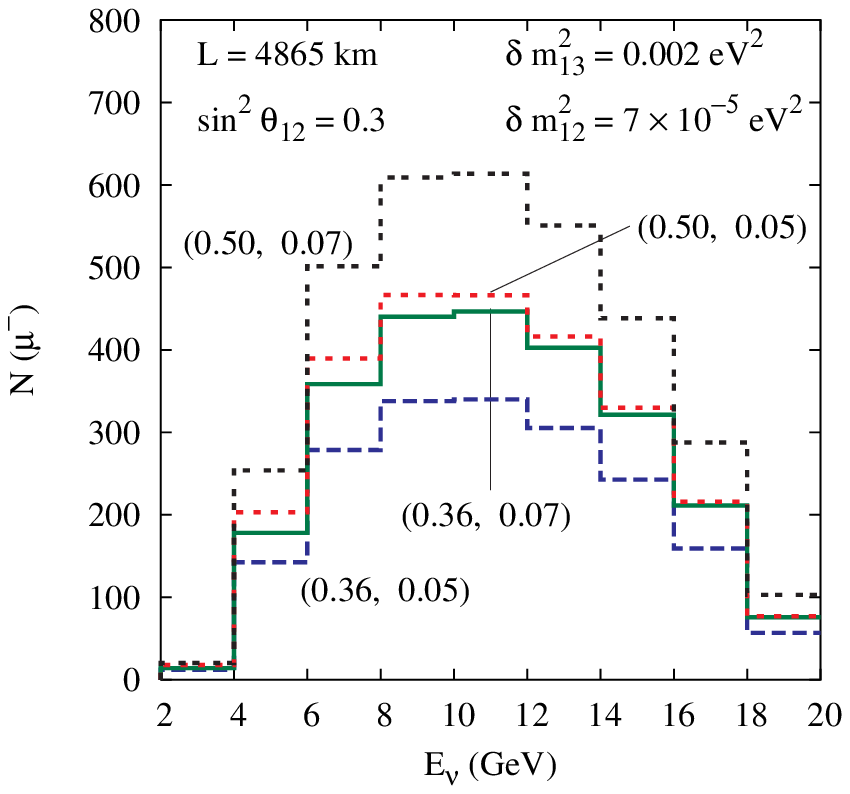}
\hspace*{-7em}
\epsfxsize = 10cm  \epsfysize = 7cm \epsfbox{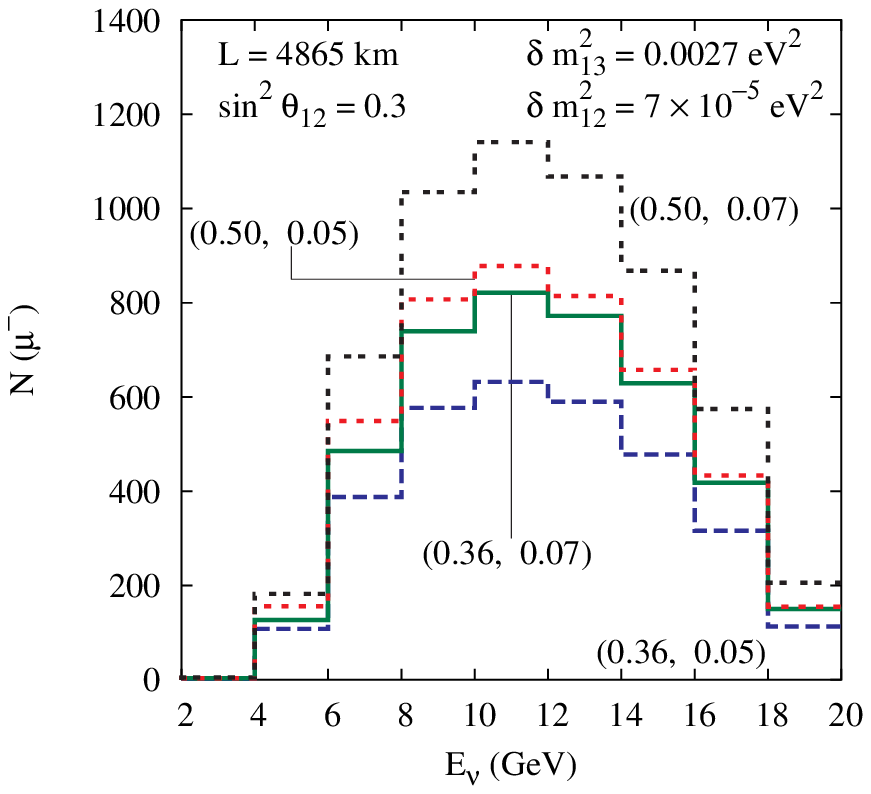}
}
\vspace*{-0.8cm}
\caption{\em The number of $\mu^-$ events in a 50 kT iron calorimeter
  exposed to a total of $2 \times 10^{20}$ $\mu^+$ decays in a
  20 GeV storage ring at a distance of 4865 km. The histograms correspond
  to different values of $(|U_{\mu 3}|^2, |U_{e 3}|^2)$ with the left
  (right) panels referring to $\delta m_{13}^2 = 0.002 \, (0.0027) \ev^2$.
}
   \label{Fig:e_mu_events}
\end{figure}

In Fig.\ref{Fig:e_mu_events}, we present the number of $\mu^-$ events in the
detector when exposed to $\mu^+$ decays. As before, the detector is supposed
to be a 50 kT one, with a total exposure of $2 \times 10^{20}$ decays. Clearly,
the number of events grows with $|U_{e3}|^2$. While the exact dependence
is not linear (as the naive expectation would be), explicit
calculations with other values of $|U_{e3}|$ shows that
it is not very far from being linear. This is also the case for
the larger baseline of 10480 km (which we do not exhibit here).
The dependence on $|U_{\mu 3}|$ is
more complicated though. As can be expected, this depends more crucially
on the size of the matter effect and hence on the exact  baseline
(see, for example, Fig.~\ref{Fig:emu_prob}). However, given the considerably
large number of such events, it is quite apparent that this measurement
is likely to be a sensitive probe in the ($|U_{\mu 3}|, |U_{e 3}|$) plane.
Unfortunately though, the event distribution is not very sensitive to
$\delta m^2_{13}$.

The main motivation for the above excercise is to show that the
measurement of any of the small parameters ($\ue$ or $\beta$) at long
baseline experiments cannot be done independent of each
other. Although experiments with relatively small baselines (such as
NuMi--{\sc minos}) seem to be quite sensitive to the parameter
$\beta$, this optimism is misplaced as explained in
Sections~\ref{sec:minos} \& \ref{sec:longbaseline}.  On the other
hand, moderate baseline experiments do show increased sensitivity to
$\beta$, but only at the cost of having a more complicated dependence
on $\beta$ and $\ue$ (see eq.\ref{survival_simp} and
Fig. \ref{Fig:comp_4865_10480}). In either case, a measurement of
$\beta$ from $\nu_\mu$-survival probability asymmetry needs an accurate
knowledge of $\ue$. It is here that the wrong-sign muon rates (which
is more sensitive to $U_{e3}$) have the most important role to
play. Of course, a precise knowledge of $\delta m^2_{13}$ is important for
the extraction of $\ue$ (although the dependence on $\delta m^2_{13}$ is
not as pronounced for the wrong-sign muon events as for the right-sign
ones). However, it is expected that the error-bars on $\delta m^2_{13}$
would be significantly reduced before the era of such neutrino factory
experiments.  Then, in principle, $A_N$ and wrong-sign muon rates
could be used simultaneously in a statistical analysis (in a spirit
similar to that followed in Ref. \cite{maoki}) to extract $\ue$ and
$\beta$ to much greater accuracy than possible today.

\section{Conclusions}
      \label{sec:concl}
Analyses of matter effects in $\nu_\mu$ oscillations have largely 
concentrated on transition probabilities. Revisting the problem, 
we demonstrate that the survival probabilities too are very sensitive 
to matter effects and can be used profitably in 
determination of several crucial parameters in the neutrino sector. 

To start with, we derive a set of approximate analytical 
expressions for the $\nu_\mu$ survival probability in presence
an arbitrarily large matter density. While similar results had 
already been reported for small matter terms in the effective 
Hamiltonian, our expressions have a much wider range of validity. 

Since we are interested in the survival probability for $\nu_\mu$, 
a variable of interest can be constructed simply from the number 
of charged current (CC) events in the detector. Starting 
with $\nu_\mu$ and $\bar \nu_\mu$ beams, 
$A_N \equiv (N_- - N_+) / (N_- + N_+)$ describes 
an asymmetry between the number of `same-sign'
$\mu^-$ and $\mu^+$ events generated in the detector after 
traversing a given baseline. We show by explicit event rate
calculation that this asymmetry is a good measure of the matter
effects felt by the neutrinos while propagating through the earth.  
However, contrary to claims in the literature~\cite{choubey_roy}, 
the size of the asymmetry for the Fermilab-{\sc minos} combine 
is much smaller than the experimental sensitivity. 

We suggest, therefore, that a future long-baseline experiment could
explore this effect. As a prototype experiment, we consider a neutrino
factory (a muon storage with muon beams of energy 20 GeV) and the
proposed 50 kT iron calorimeter detector~\cite{mono,ino} with a
capability of muon charge determination. Using realistic experimental
resolutions, we demonstrate that such a combination is sufficient to
establish the aforementioned asymmetry with a very large statistical
significance. However, unlike in the case of relatively short-baseline
Fermilab--{\sc minos} combine, the asymmetry for a long baseline is no
longer proportional to the deviation of $\Um^2$ from maximality. Even
so, the sensitivity to $(1 - 2 \, \Um^2)$ remains quite pronounced for
baselines of upto about 6500 Km. Using the aforementioned approximate
expressions for the survival probabilities, such data can thus be used
to determine $(1 - 2 \, \Um^2)$. While the sensitivity to this
combination decreases for much longer baselines, asymmetry data for
such baselines can still be used for precise determination of $\ue$.

The latter exercise is also shown to be aided by the 
measurement of `wrong-sign-muon' events generated through 
$\nu_e \to \nu_\mu$ oscillations.  
Estimating the rate of such events for the 
same experimental setup, we demonstrate that this 
could be used in conjunction with the asymmetry measurement 
to lead to a good simultaneous determination of $\Um$ and $\ue$.

\section*{Acknowledgements}
The authors thank the Theory Division, CERN for hospitality during the
initiation phase of the project. DC thanks INFN, Sezione di Roma, for
hospitality during which a major part of this work has been done. DC
acknowledges financial assistance under the {\em Swarnajayanti
Fellowship} grant from the Department of Science and Technology,
India. AD is partially supported by the RTN European Programme
MRTN-CT-2004-503369 (Quest for Unification).

\newcommand{\plb}[3]{{Phys. Lett.} {\bf B#1}, #2 (#3)}                  %
\newcommand{\prl}[3]{Phys. Rev. Lett. {\bf #1}, #2 (#3) }        %
\newcommand{\rmp}[3]{Rev. Mod.  Phys. {\bf #1} #2 (#3)}             %
\newcommand{\prep}[3]{Phys. Rep. {\bf #1} #2 (#3)}                   %
\newcommand{\rpp}[3]{Rep. Prog. Phys. {\bf #1} #2 (#3)}             %
\newcommand{\prd}[3]{Phys. Rev. {\bf D#1}, #2 (#3)}                    %
\newcommand{\np}[3]{Nucl. Phys. {\bf B#1} #2 (#3)}                     %
\newcommand{\npbps}[3]{Nucl. Phys. B (Proc. Suppl.)
           {\bf #1} #2 (#3)} %
\newcommand{\sci}[3]{Science {\bf #1} #2 (#3)}                 %
\newcommand{\zp}[3]{Z.~Phys. C{\bf#1} #2 (#3)}
\newcommand{\epj}[3]{Eur. Phys. J. {\bf C#1} #2 (#3)}
\newcommand{\mpla}[3]{Mod. Phys. Lett. {\bf A#1} #2 (#3)}             %
 \newcommand{\apj}[3]{ Astrophys. J.\/ {\bf #1} #2 (#3)}       %
\newcommand{\jhep}[2]{{Jour. High Energy Phys.\/} {\bf #1} (#2) }%
\newcommand{\astropp}[3]{Astropart. Phys. {\bf #1}, #2 (#3)}            %
\newcommand{\ib}[3]{{ ibid.\/} {\bf #1} #2 (#3)}                    %
\newcommand{\nat}[3]{Nature (London) {\bf #1} #2 (#3)}         %
 \newcommand{\app}[3]{{ Acta Phys. Polon.   B\/}{\bf #1} #2 (#3)}%
\newcommand{\nuovocim}[3]{Nuovo Cim. {\bf C#1} #2 (#3)}         %
\newcommand{\yadfiz}[4]{Yad. Fiz. {\bf #1} #2 (#3);             %
Sov. J. Nucl.  Phys. {\bf #1} #3 (#4)]}               %
\newcommand{\jetp}[6]{{Zh. Eksp. Teor. Fiz.\/} {\bf #1} (#3) #2;
           {JETP } {\bf #4} (#6) #5}%
\newcommand{\philt}[3]{Phil. Trans. Roy. Soc. London A {\bf #1} #2
        (#3)}                                                          %
\newcommand{\hepph}[1]{hep--ph/#1}           %
\newcommand{\hepex}[1]{hep--ex/#1}           %
\newcommand{\astro}[1]{(astro--ph/#1)}         %


\begin{thebibliography}{99}
\bibitem{superK}
Super-Kamiokande collaboration, Y. Hayato, talk given at the EPS 2003
conference (Aachen, Germany, 2003),
{\tt http://eps2003.physik.rwth-aachen.de};\\
 S. Fukuda et al., Phys. Lett. {\bf B539}, 179 (2002).
\bibitem{sno}
SNO collaboration, S.N. Ahmed et al., nucl-ex/0309004.
\bibitem{chooz}
CHOOZ collaboration, M. Appolonio et al., Eur. Phys. J {\bf C27}, 331
(2003).

\bibitem{kamland}
KamLAND collaboration, K. Eguchi et al., Phys. Rev. Lett. {\bf 90},
021802 (2003).

\bibitem{k2k}
K2K collaboration, I. Kato, talk given at the 38th {\it Recontres de
Moriond on Electroweak Interactions and Unified Theories} (Les Ares,
France, 2003), hep-ex/0306043.

\bibitem{review}For a recent review for the limits on masses and mixing, see
S. Goswami, A. Bandyopadhyay and  S. Choubey, hep-ph/0409224.

\bibitem{msw}L. Wolfenstein, \prd{17}{2369}{1978}; \\
S. Mikheyev and A. Smirnov, Sov. J. Nucl. Phys. {\bf 42}, 913 (1985).

\bibitem{nu_fac_tech}For example, see S. Geer, FERMILAB-CONF-04-133-E, 
2004;  \\
R. Raja,. FERMILAB-CONF-04-018-E (2004), hep-ex/0402022;\\
A. Blondel et. al., Nucl.Instrum. Meth. {\bf A451},102 (2000).

\bibitem{nu_fac}V. Barger, S. Geer, R. Raja and K. Whisnanat, 
\prd{62}{013004}{2000}; \\
  C. Albright et al., \hepex{0008064}; \\
M. Freund, P. Huber and M. Lindner, \np{585}{105}{2000}; \np{615}{331}{2001};\\
M. Apollonio et al., \hepph{0210192}.

\bibitem{barger_matter}V. Barger, S. Geer, R. Raja and K. Whisnanat,
Phys.Lett. {\bf B485} 379 (2000); \\
A. Cervera et al., \np{579}{17}{2000}.

\bibitem{maoki}M. Aoki et al., \prd{67}{093004}{2003}; \\
M. Aoki, K, Hagiwara and N. Okamura, \hepph{0311324}.

\bibitem{raj} R. Gandhi et al., hep-ph/0408361.

\bibitem{mono}N.Y.~Agafonova {\it et al.}  [MONOLITH Collaboration],
LNGS-P26-2000, LNGS-P26-00, CERN-SPSC-2000-031, CERN-SPSC-M-657; see
{\tt http://castore.mi.infn.it/$\sim$monolith/}

\bibitem{ino}See
{\tt http://www.imsc.res.in/$\sim$ino}; and working reports and talks
therein.

\bibitem{choubey_roy}
S. Choubey and P. Roy, Phys. Rev. Lett. {\bf 93}, 021803  (2004).

\bibitem{prem}
A. M. Dziewonski and D.L. Anderson, 
Phys. Earth Plan. Int. {\bf 25}, 297 (1981); \\
we use the
parametrisation given in R. Gandhi, C. Quigg, M. Reno and I. Sarcevic,
\astropp{5}{81}{1996}.

\bibitem{expand}E. Akhmedov et al., \jhep{0404:078}{2004}.

\bibitem{minos}
R. Saakian et al ({\sc minos} collaboration), 
      Nucl. Phys. Proc. Suppl. {\bf 111}, 169 (2002); Yad.Fiz. {\bf 67}, 
      1112 (2004); \\
 M.V. Diwan et al.,  Nucl. Phys. Proc. Suppl. {\bf 123}, 272 (2003).

%
\bibitem{minerva}MINERvA Collaboration (D.A. Harris for the collab.),
    FERMILAB-PUB-04-252-E, hep-ex/0410005;

\bibitem{jhf}Y. Itow et al., hep-ex/0106019; also see :
{\tt http://neutrino.kek.jp/jhfnu}.


\end{thebibliography}
\end{document}